\RequirePackage{ifpdf}
\ifpdf
  \documentclass[10pt,a4paper,pdftex]{article}
\else
  \documentclass[10pt,a4paper,dvips]{article}
\fi

\ifpdf
  \usepackage{color}
\fi

\usepackage[latin1]{inputenc}
\usepackage{amsmath}
\usepackage{amsfonts}
\usepackage{amssymb}
\usepackage{float}
\usepackage{epsfig}
\usepackage{subfigure}
\usepackage[section]{placeins}
\usepackage{pstricks}
\usepackage{mathrsfs}

\ifpdf
  \usepackage{graphicx}
  \usepackage[unicode]{hyperref}
\else
  \usepackage{graphics}
\fi

\newcommand{\linkFixer}[2]{\ifpdf
			     \texorpdfstring{#1}{#2}
			   \else
			     #1
			   \fi
			  }

\newcommand{\td}{\mathrm{d}}

\newcommand{\order}[1]{\mathcal{O}\left(#1\right)}

\newcommand{\Lie}[1]{{#1}}

\newcommand{\LevCi}[3]{\left\{ \begin{smallmatrix} #1 \\ #2
      #3 \end{smallmatrix}\right\} }

\allowdisplaybreaks[1]

\author{Jérôme Gaillard\footnote{pyjg@swansea.ac.uk} \; and \:
  Johannes Schmude\footnote{pyjs@swansea.ac.uk}\\
  \mbox{}\\
  Department of Physics\\
  Swansea University, Swansea, SA2 8PP, United Kingdom}

\title{The lift of type IIA supergravity with D6 sources:
  M-theory with torsion}

\date{}

\begin{document}
\numberwithin{equation}{section}

\maketitle

\begin{abstract}
  This paper is concerned with an extension of the well known
  Kaluza-Klein mechanism. As the standard ansatz for Kaluza-Klein
  reduction implies the existence of a gauge potential associated with
  the KK field 
  strength, it follows immediately that this field strength
  satisfies its Bianchi identity. Hence, the standard KK formalism
  breaks down in the presence of a violated Bianchi identity. This
  occurs for example in the context of D6 sources.

  We will investigate and partially solve this problem 
  in the context of the type IIA/M-theory duality. Our discussion is
  motivated by the construction of
  gauge/string duals with backreacting flavor branes using D6-branes,
  which appear in M-theory as KK-monopoles.

  We are able to derive
  source-modified equations of motion for the
  eleven-dimensional theory, and are subsequently able to
  obtain the source-modified type IIA equations by direct
  dimensional reduction.
\end{abstract}

\newpage

\tableofcontents

\section{Introduction}
\label{sec:introduction}

In the context of M-theory, the relations between type IIA string
theory and eleven-dimensional supergravity are by now standard
textbook material (see for example
\cite{Kiritsis:2007zz,Becker:2007zj,Ortin:2004ms,Johnson:2003gi}). The
M2-brane gives rise to the D2
and the fundamental string, the M5 to the D4 and NS5 branes. The D0
and D6-branes on the other hand have a 
slightly different origin. Not being related to any brane-like object
in eleven dimensions, they are results of the Kaluza-Klein (KK)
reduction relating the two theories; the former being a particle-like,
localized gravitational excitation on the KK-circle, the latter a
peculiar fibration of said circle over the ten-dimensional base, known
as a Kaluza-Klein monopole (a good review is given by
\cite{Gubser:2002mz}). In this paper, we are concerned with a
small gap in this formalism that becomes apparent when one tries
to consider the M-theory lift of smeared D6-branes.

The problem can be quickly explained. The bosonic sector of
eleven-dimen\-sional supergravity contains only the graviton
$\hat{g}_{MN}$ and a four-form field $\hat{F}_{(4)}$. Upon KK
reduction, $\hat{F}_{(4)}$ gives rise to the Kalb-Ramond three-form
field $H_{(3)}$ as well as the Ramond-Ramond four-form $F_{(4)}$. From
$\hat{g}_{MN}$ one obtains the ten-dimensional metric $g_{\mu \nu}$,
the dilaton $\Phi$, and a one-form gauge potential $A_{(1)}$, with an
associated field strength $F_{(2)} = \td A_{(1)}$. If we assume the
KK-circle to be parameterized by $z$, the standard KK-ansatz relating
the two geometries is\footnote{Where the distinction is necessary,
  hats and tildes denote eleven-dimensional quantities. Capital
  letters describe eleven-dimensional indices. The M-theory circle
  will be parameterized by either $z$, $\psi_+$ or $\psi$.
}
\begin{equation}\label{eq:Standard_KK-ansatz}
  \begin{aligned}
    \td s_{\text{M}}^2 &= e^{-\frac{2}{3}\Phi} \td s_{\text{IIA}}^2 +
    e^{\frac{4}{3}\Phi} ( A_{(1)} + \td z )^2 \\
    \hat{F}_{(4)} &= F_{(4)} + H_{(3)} \wedge \td z
  \end{aligned}
\end{equation}
Given any solution to the equations of motion of type IIA
supergravity, one can use (\ref{eq:Standard_KK-ansatz}) to lift to
eleven dimensions and vice versa. However, as $A_{(1)}$ plays the role
of a gauge potential, it is actually $F_{(2)} = \td A_{(1)}$ that
contains the physically relevant degrees of freedom. Thus given a set
$\{ g_{\mu\nu}, \Phi, F_{(2)}, H_{(3)}, F_{(4)} \}$ one first has to
find a gauge potential prior to lifting. Now assume that for some reason $\td F_{(2)} \neq 0$. Clearly $A_{(1)}$ cannot exist
and we are unable to find a gauge potential. Therefore we cannot use
(\ref{eq:Standard_KK-ansatz}) to perform the lift. This is the
apparent gap in the standard formalism we alluded to earlier.

The problem is not a purely formal
one. D6-branes couple magnetically to $A_{(1)}$. As we will explain
shortly, the inclusion of D6 sources violates the Bianchi identity
$\td F_{(2)} = 0$ at the position of the sources. While this is not a
problem for localized sources -- as a matter of fact it is the reason
why the KK-monopole is a gravitational instanton -- one encounters the
problem at hand once one distributes the branes continuously and thus
violates the Bianchi identity on an open subset of space-time.

As an aside it is worthwhile to point out that the relation between D6-branes and the RR two-form is much the same as that between magnetic
monopoles and the $F_{\text{E\&M}}$ in standard
electro-magnetism. The inclusion of magnetic sources restores the
symmetry of the Maxwell equations. Schematically
\begin{equation}
  \begin{aligned}
    \td * F_{\text{E\&M}} &= *j_{\text{E}} & 
    &\quad &
    \td F_{\text{E\&M}} &= *j_{\text{M}}
  \end{aligned}
\end{equation}
Thus, the Bianchi identity is violated by the magnetic current
$j_{\text{M}}$. In the context of quantum field theories one speaks
of monopole condensation. (See e.g.~\cite{Quevedo:1996uu})

In this paper, we will not resolve the issue in full generality, but
will focus on the inclusion of D6 sources in type IIA backgrounds of
the form
\begin{equation}
  \mathcal{M}_{10} = \mathbb{R}^{1,3} \times \mathcal{M}_6
\end{equation}
without three of four-form flux, that preserve four supercharges. More
precisely, we will be interested in the construction of string duals
to $3+1$-dimensional $\Lie{SU}(N_c)$ gauge theories with $\mathcal{N}
= 1$ supersymmetry and $N_f$ flavors using D6-branes. Let us briefly
expose some general points concerning gauge/string dualities.

Since its discovery, the AdS/CFT correspondence (\cite{Maldacena:1997re}, \cite{Witten:1998qj}) has been used to study
a variety of problems ranging from black hole entropy to the physics
of condensed matter systems. In the line of work dedicated to the
study of increasingly realistic gauge theories with physics similar to
the one of QCD, recent years have seen a continuous extension of the
duality to gauge theories with matter fields charged under the
fundamental representation of the gauge group \cite{Karch:2002sh}. On
the string theory
side, the addition of fundamental matter corresponds to the
inclusion of branes extending along both the Minkowski directions
associated with the gauge theory as well as a non-compact cycle
transverse to them. As long as one remains in the proble limit, $N_f
\ll N_c$, it is sufficient to consider these flavor branes as probes
in the background space-time. If one wants to go
beyond this probe approximation (\cite{Bigazzi:2005md,Casero:2006pt,Casero:2007jj,HoyosBadajoz:2008fw}),
one needs to include the backreaction
of the flavor branes onto space-time. I.e.~one needs to consider the
combined action
\begin{equation}\label{eq:schematic_flavor-modified_action}
  S = S_{IIA} + S_{\text{Branes}}
\end{equation}
where $S_{\text{Branes}} = \sum_{N_f} ( S_{DBI} + S_{WZ} )$ consists
of the standard brane action for every single flavor brane. For other
examples, see
\cite{Benini:2006hh,Caceres:2007mu,Benini:2007gx,Casero:2007pz,Benini:2007kg,Bigazzi:2008zt,Canoura:2008at,Arean:2008az,Paredes:2006wb,Zeng:2007ta,Bigazzi:2008gd,Bigazzi:2008ie,Bigazzi:2008qq}.

Note that apart from \cite{Karch:2002sh}, all the previously cited
duals consider the addition of flavors to string duals with reduced
supersymmetry. One of the main techniques used to reduce the amount of
supercharges preserved by the geometry consists in wrapping
the $N_c$ color branes on compact cycles in the geometry
\cite{Witten:1998zw,Maldacena:2000yy,Gomis:2001aa,Gauntlett:2001ps,Bigazzi:2001aj}.
E.g.~one wraps D4-branes on an $S^1$ or D5s on an $S^2$ to study $3+1$-dimensional gauge theories. In general, the preservation of some
supersymmetry requires the gauge theory living on the brane to
be topologically twisted -- see \cite{Maldacena:2000yy}.

This paper was originally born out of the interest to study the
addition of flavor branes to type IIA backgrounds dual to
$\mathcal{N}=1$, $\Lie{SU}(N_c)$ super Yang-Mills. Before flavoring,
the geometry is that of $N_c$ D6-branes wrapping a three-cycle in the
deformed conifold.\footnote{To be precise, we will be dealing with conifolds deformed by the presence of branes or $F_{(2)}$ flux. They do carry $\Lie{SU}(3)$-structure but are not of $\Lie{SU}(3)$-holonomy. Therefore, they are not Calabi-Yau and strictly speaking we should not refer to them as (deformed/resolved) conifolds. For the lack of a better term however, we shall refer to the internal six-dimensional manifolds in this paper by that name though, as their topology is the same as that of their Calabi-Yau cousins.}
In the limit $N_c g_{\text{YM}}^2 \gg 1$, the backreaction of the
color branes causes the system to undergo a geometric transition. The
system is now best described in terms of the resolved conifold with
the branes having been replaced by $N_c$
units of two-form flux over a two-cycle. This was originally
studied in \cite{Atiyah:2000zz,Edelstein:2001pu} and the geometric
transition is based on the work of
\cite{Gopakumar:1998ki,Vafa:2000wi}; an attempt at generalizing the
duality to include finite-temperature duals was made in
\cite{Schmude:2007dg}. The resulting ten-dimensional 
background consists of metric, dilaton and RR two-form $(g_{\mu\nu},
\Phi, F_{(2)})$. Refering back to (\ref{eq:Standard_KK-ansatz}) one
sees that it lifts to pure geometry in M-theory, as both $H_{(3)}$ and
$F_{(4)}$ are set to zero. It is for this reason that it is
particularly simple and interesting to study these geometries and
dualities from the perspective of eleven-dimensional
supergravity. Here, the equations of motion and supergravity
variations simplify to
\begin{equation}
  \label{eq:M-theory_eom_and_SUSY-variations_pure_gravity}
  \begin{aligned}
    \hat{R}_{MN} &= 0 &
    &\quad &
    \delta_{\hat{\epsilon}} \hat{\psi}_M &= \partial_M \hat{\epsilon}
    + \frac{1}{4} \hat{\omega}_{M A B} \hat{\Gamma}^{A B}
    \hat{\epsilon}
  \end{aligned}
\end{equation}
The eleven-dimensional geometry is of the form
\begin{equation}
  \mathcal{M}_{11} = \mathbb{R}^{1,3} \times \mathcal{M}_7  
\end{equation}
As the seven-dimensional manifold $\mathcal{M}_7$ preserves 1/8-SUSY and is Ricci
flat, it is a manifold of $G_2$-holonomy. The concept of M-theory
compactifications on such manifolds (\cite{Atiyah:2001qf}) is pretty
much the same as that of the old heterotic string models on Calabi-Yau
three-folds used in classic string phenomenology. Mathematically this
is reflected by the presence of a three-form $\hat{\phi}_{G_2}$ that
is closed and co-closed
\begin{equation}\label{eq:G2-structure_with_G2-holonomy}
  \begin{aligned}
    \td \hat{\phi}_{G_2} &= 0 &
    &\quad &
    \td ( *_7 \hat{\phi}_{G_2} ) &= 0
  \end{aligned}
\end{equation}
where $*_7$ denotes the seven-dimensional Hodge dual on the internal
space.

From the point of view of type IIA string theory, the flavoring
procedure is reasonably straightforward. As was shown
in \cite{Gutowski:1999tu,Koerber:2005qi,Koerber:2006hh} and then applied to
gauge/string duality in \cite{Gaillard:2008wt}, the brane action can
be written as an integral over the ten-dimensional space-time instead of as a
sum over integrals over the seven-dimensional world volume,
\begin{equation}
  \begin{aligned}
    S_{\text{Branes}} &= -T_6 \int_{\mathcal{M}_{10}} (e^{-\Phi} \phi_{D6} - A_{(7)}
    ) \wedge \Xi_{(3)}
  \end{aligned}
\end{equation}
where $\phi_{D6}$ is the so-called calibration form and $\Xi_{(3)}$
takes the role of a source density for the D6-branes. The presence of
$S_{\text{Branes}}$ in the modified action
(\ref{eq:schematic_flavor-modified_action}) gives source term
contributions to the equations of motion. Most prominent among these
is the appearance of a magnetic source term for the RR two-form,
\begin{equation}\label{eq:violation_of_bianchi_identity}
  \begin{aligned}
    \td F_{(2)} &= - (2 \kappa_{10}^2 T_6) \Xi_{(3)}
  \end{aligned}
\end{equation}
that violates the standard Bianchi identity. In type IIA one
accomodates for this simply by adding a flavor contribution to the RR
form,
\begin{equation}
  \begin{aligned}
    F_{(2)} &= \td A_{(1)} + (2 \kappa_{10}^2 T_6) B_{(2)}
  \end{aligned}
\end{equation}
with $B_{(2)} \to 0$ as $N_f \to 0$. (Note that $B_{(2)}$ is not to be
confused with the Kalb-Ramond two-form potential $H_{(3)} = \td B_{(2)}$
that will not appear in this paper.) On a technical
side, one anticipates that the flavor branes will deform the $N_f = 0$
geometry, and begins therefore by studying deformations of the
original background prior to flavoring. Subsequently one searches for
solutions of the fully backreacted problem. Intuitively one
would consider a localized stack of $N_f$ flavor branes from which it
follows that $\Xi_{(3)}$ should contain delta functions localizing the
sources on the internal cycles. This is undesirable as localized
sources break the isometries of the background and do therefore break
global symmetries of the full dual theory (including the KK-modes); the resulting differential
equations are also very hard to solve. Therefore, one distributes the
flavor branes continuously over their transverse cycles. In the
process, the flavor symmetry breaks as $\Lie{U}(N_f) \mapsto
\Lie{U}(1)^{N_f}$. The procedure is known as smearing and the form
$\Xi_{(3)}$ occasionally refered to as the smearing form. As it was
shown in \cite{Gaillard:2008wt}, the choice of smearing form is not
arbitrary as supersymmetry and the modified Bianchi identities require
it to satisfy
\begin{equation}
  \begin{aligned}
    \td *_{10} \td (e^{-\Phi}\phi_{D6}) = -(2 \kappa_{10}^2 T_6) \Xi_{(3)}
  \end{aligned}
\end{equation}

It is a priori not obvious how to accomodate the violation of the
Bianchi identity (\ref{eq:violation_of_bianchi_identity}) in
M-theory. However, as the sources will not only modify the Bianchi
identity, yet also the dilaton and Einstein equations, it is
reasonable to expect that the eleven-dimensional geometry will not be
Ricci flat. Instead, the Einstein equations should be supplemented by
the presence of a source term,
\begin{equation} \label{eq:Einstein_equation}
  \begin{aligned}
    \hat{R}_{M N} - \frac{1}{2} \hat{g}_{MN} \hat{R} &= \hat{T}_{M N}
  \end{aligned}
\end{equation}
From the loss of Ricci flatness it follows that the manifold can no
longer be of $G_2$-holonomy; as it preserves the same amount of
supersymmetry however it is fair to expect it to carry a $G_2$-structure. Therefore, there is still a three-form $\hat{\phi}_{G_2}$
that now fails to be (co)closed. One can anticipate that the failure
of the manifold to be of $G_2$-holonomy is parameterized by $N_f$ and
thus ultimately by the $B_{(2)}$ contribution to $F_{(2)}$, i.e.
\begin{equation}
  \begin{aligned}
    \td \hat{\phi}_{G_2} &\sim (F - \td A) \\
    \td (*_7\hat{\phi}_{G_2}) &\sim (F - \td A)
  \end{aligned}
\end{equation}
These expressions, relating forms of different degrees, are to be
understood in such a way that the left hand side vanishes when the right
hand side does, and vice versa. Now for a manifold carrying a
$G$-structure, its failure to be of $G$-holonomy is measured by its
intrinsic torsion.\footnote{For intrinsic
  torsion in the context of string theory see
  \cite{Gauntlett:2003cy}.
}
Therefore, we expect
the flavors in eleven dimensions to appear in the form of intrinsic
torsion. A detailed study of the relation between the eleven and
ten-dimensional supersymmetry variations will prompt us to consider
eleven-dimensional backgrounds with torsion $\hat{\tau}$, where the
torsion is related to $F - \td A = B$.

Finally we will see that an uplift of our ten-dimensional equations of
motion is given by the relation
\begin{equation}\label{eq:Source-modified_m-thy_eom}
  \begin{aligned}
    R^{(\tau)}_{MN} + \frac{1}{2}
    R^{(\tau)}_{K L R N} (*_7
    \hat{\phi})_{M}^{\phantom{M}K L R} = 0
  \end{aligned}
\end{equation}
which is the solution to our initial problem. $R^{(\tau)}$ is the
eleven-dimensional Riemann (Ricci) tensor with torsion -- we have
discarded the use of hats to avoid an overly cluttered notation.
As one can always rewrite the Riemann tensor as a combination of a
torsion free Riemann tensor with additional terms depending on the
torsion, it is possible to recast the above equation in the form of
(\ref{eq:Einstein_equation}) with the energy-momentum tensor depending
only on the torsion.

At first glance, equation (\ref{eq:Source-modified_m-thy_eom}) appears
like a modification of M-theory and violates all 
intuition as eleven-dimensional supergravity is unique.
However, one must not forget that we never assumed to solve
the problem in its full generality. As a matter of fact,
(\ref{eq:Source-modified_m-thy_eom}) has to be taken with several
pinches of salt -- which might not be a surprise, as the inclusion of
source terms in theories of gravity is always a rather difficult
business. First of all, (\ref{eq:Source-modified_m-thy_eom}) assumes
the background to be of topology $\mathcal{M}_{11} = \mathbb{R}^{1,3}
\times \mathcal{M}_7$, with the internal manifold carrying a $G_2$-structure. Furthermore this means that we are not dealing with maximal
eleven-dimensional supergravity, but with a situation with reduced 
supersymmetry -- 1/8 BPS -- in which case the theory is no longer
unique. Still, as we will see, equation
(\ref{eq:Source-modified_m-thy_eom}) manages what the standard KK-ansatz (\ref{eq:Standard_KK-ansatz}) does not. It gives the correct
source-modified equations of motion in type IIA.

The structure of this paper is as follows. In section
\ref{sec:flavoring-iia} we will
begin with a review of the unflavored geometries in ten and eleven
dimensions and then continue by studying the flavoring problem from
the perspective of type IIA. Following this, we will turn to the issue
of the M-theory lift in section \ref{sec:back-11d}. The paper is
ammended by several appendices on brane embeddings, spinor conventions
and KK reduction. For illustrative and motivational purposes we will
be using a specific case of a M-theory $G_2$-holonomy manifold and its
type IIA reduction in section \ref{sec:flavoring-iia}. However, the
results of section \ref{sec:back-11d} on the M-theory lift of smeared
D6-branes do not depend on this example or the type IIA reduction
chosen. They only depend on the presence of a $G_2$-structure, four-dimensional Minkowski space and the absence of M-theory fluxes. 

Note that (\ref{eq:Source-modified_m-thy_eom}) is not the only result
presented here. As we are studying the flavoring problem in type IIA in
order to find an answer to the issue of the M-theory lift, this paper
makes also considerable progress towards the construction of a
dual to four-dimensional, $\mathcal{N}=1$ $\Lie{SU}(N_c)$
super Yang-Mills with backreacting flavors. For the specific ansatz of
section \ref{sec:flavoring-iia}, we are able to derive a set of very
generic first-order equations -- (\ref{eq:10D_BPS}) and
(\ref{eq:F_Derivatives_Conditions}) -- that have to be satisfied by
smeared D6 sources in this geometry. We proceed to derive an analytic
one-parameter family of solutions in section
\ref{sec:discussing_the_iia_solution}. While the fluxes in this
solution satisfy the flux quantization necessary for a string dual,
the geometry is that of a cone over $S^2 \times S^3$ with a
singularity at the origin. So we expect the interpretation of this
solution as a suitable dual to be difficult. The presentation of the
flavoring problem is supplemented by a discussion of D6-brane
embeddings for the geometries at hand in appendix
\ref{sec:brane_embeddings}.

\section{\linkFixer{Flavored $\mathcal{N}=1$ string
    duals from D6-branes}{Flavored N=1 string duals from D6-branes}}
\label{sec:flavoring-iia}

In this section, we will review the source-free string duals in their
ten and eleven-dimensional formulations. Subsequently we will be turning to
the issue of adding sources to the type IIA background. Let us once
more emphasize that the particular choices of eleven-dimensional
geometry (and its dimensional reduction) are of no direct
consequence for our results concerning the M-theory lift of smeared
D6-branes. The concrete geometry presented here is chosen due to its
relevance to the flavoring problem in type IIA.

\subsection{The eleven-dimensional dual without sources}
\label{sec:brandhubers-paper}
Building on the work of Brandhuber \cite{Brandhuber:2001kq} (see also
\cite{Brandhuber:2001yi,Cvetic:2001kp}) we consider the purely gravitational
M-theory background given by the elfbein
\begin{equation}
  \label{eq:d=11_unflavored_elfbein_(brandhuber)}
  \begin{aligned}
    \tilde{e}^\mu &= \td x^\mu &
    \tilde{e}^\rho &= E(\rho) \td\rho \\
    \tilde{e}^{1,2} &= A(\rho) \sigma_{1,2} &
    \tilde{e}^{3,4} &= C(\rho) \lbrack \Sigma_{1,2} - f(\rho)
    \sigma_{1,2}\rbrack \\
    \tilde{e}^5 &= B(\rho) \sigma_3 &
    \tilde{e}^6 &= D(\rho) \lbrack \Sigma_3 - g(\rho) \sigma_3 \rbrack
  \end{aligned}
\end{equation}
$\sigma_i, \Sigma_i$ are left-invariant Maurer-Cartan forms which we
chose to be
\begin{equation}
  \label{eq:Maurer-Cartan-forms}
  \begin{aligned}
    \sigma_1 &= \cos\psi \td\theta + \sin\psi \sin\theta \td\phi &
    \Sigma_1 &= \cos\tilde{\psi} \td\tilde{\theta} + \sin\tilde{\psi}
    \sin\tilde{\theta} \td\tilde{\phi} \\
    \sigma_2 &= -\sin\psi \td\theta + \cos\psi \sin\theta \td\phi &
    \Sigma_2 &= -\sin\tilde{\psi} \td\tilde{\theta} + \cos\tilde{\psi}
    \sin\tilde{\theta} \td\tilde{\phi} \\
    \sigma_3 &= \td\psi + \cos\theta \td\phi &
    \Sigma_3 &= \td\tilde{\psi} + \cos\tilde{\theta}
  \end{aligned}
\end{equation}
The solutions we are interested in are 1/8-BPS; therefore one can
impose the following constraints onto the SUSY spinor
$\tilde{\epsilon}$:
\begin{equation}
  \label{eq:m-thy_SUSY-spinor_non-reducible-gauge}
  \begin{aligned}
    \tilde{\Gamma}^{1234} \tilde{\epsilon} &= \tilde{\epsilon} &
    \tilde{\Gamma}^{1356} \tilde{\epsilon} &= -\tilde{\epsilon} &
    \tilde{\Gamma}^{\rho 126} \tilde{\epsilon} &= -\tilde{\epsilon}
  \end{aligned}
\end{equation}
As a direct consequence we can calculate the
following spinor bilinear, which turns out to be the $G_2$-structure form
\begin{equation}
  \label{eq:m-thy_G2-structure_no-flavors}
  \begin{aligned}
    \tilde{\phi}_{G_2} &= (\bar{\tilde{\epsilon}} \tilde{\Gamma}_{A_0
      A_1 A_2} \tilde{\epsilon}) \tilde{e}^{A_0 A_1 A_2} \\
    &= \tilde{e}^{\rho 13} + \tilde{e}^{\rho 24} + \tilde{e}^{\rho 56}
    + \tilde{e}^{146} + \tilde{e}^{345} - \tilde{e}^{125} - \tilde{e}^{236}
  \end{aligned}
\end{equation}
In the absence of four-form flux the preservation of four supercharges
is equivalent to the manifold being of $G_2$-holonomy. A necessary and
sufficient condition is the closure and co-closure of the $G_2$-structure form. By imposing $\td \tilde{\phi}_{G_2} = 0$ and $\td (*_7
\tilde{\phi}_{G_2}) = 0$ we obtain the BPS equations
\begin{equation}
  \label{eq:m-thy_bps-equations_no-flavors}
  \begin{aligned}
    A^\prime &= \frac{E \lbrack B D (g - f^2) + A C f (1-g)\rbrack}{2
      AB} &
    B^\prime &= \frac{EC f (1-g)}{A} \\
    D^\prime &= \frac{E \lbrack A^2 ( 2 C^2 - D^2 ) + C^2 D^2 (f^2 -
      g) \rbrack}{2 A^2 C^2} &
    C^\prime &= \frac{E \lbrack A B D - C^3 f (1-g)\rbrack}{2 A B C} \\
    f &= \frac{B C}{2 A D} &
    g &= 1 - 2f^2
  \end{aligned}
\end{equation}
The same BPS system follows from demanding that
$\delta_{\tilde{\epsilon}} \tilde{\psi}_M = 0$.

The best known solution to (\ref{eq:m-thy_bps-equations_no-flavors})
is the Bryant-Salamon metric \cite{Bryand:1989mv}. With
\begin{equation}
  \label{eq:Bryant-Salamon_solution}
  \begin{aligned}
    A^2 &= B^2 = \frac{\rho^2}{12} &
    C^2 &= D^2 = \frac{\rho^2}{9} (1 - \frac{\rho_0^3}{\rho^3}) &
    E^2 &= (1-\frac{\rho_0^3}{\rho^3})^{-1} &
    f &= g = \frac{1}{2}
  \end{aligned}
\end{equation}
the metric takes the form
\begin{equation}
  \label{eq:Bryant-Salamon_metric}
  \begin{aligned}
    \td s^2 &= \td x_{1,3}^2 + (1-\frac{\rho_0^3}{\rho^3})^{-1}
    \td\rho^2 + \frac{\rho^2}{12} \sigma^2 + \frac{\rho^2}{9}
    (1-\frac{\rho_0^3}{\rho^3}) (\Sigma - \frac{1}{2}\sigma)^2
  \end{aligned}
\end{equation}
The seven-dimensional $G_2$ cone actually turns out to be the
cotangent bundle $T^* S^3$. The geometry is that of a cone over $S^3
\times S^3$, with each sphere being parameterized by a set of
Maurer-Cartan forms. At $\rho = \rho_0$, the minimum of the radial
parameter, one of the spheres ($\Sigma$) collapses, while the other
($\sigma$) remains of finite size. M-theory dynamics on this type of
manifold were discussed in \cite{Atiyah:2001qf}. Fluctuations in
$\rho_0$ and the gauge potential $A_3$ can be combined into a complex
parameter. However, as these fluctuations turn out to be
non-normalizable, they do not parameterize a moduli space of vacua, yet
rather a moduli space of theories.

There are three $\Lie{U}(1)$ isometries in
(\ref{eq:d=11_unflavored_elfbein_(brandhuber)}) given by
$\partial_\phi$, $\partial_{\tilde{\phi}}$ and
$\partial_\psi + \partial_{\tilde{\psi}}$ and there are therefore
three different dimensional reductions to type IIA. In each case one
obtains a conifold geometry with flux, with the conifold singularity
being resolved by a deformation or resolution. I.e.~there is a cone over $S^2
\times S^3$ and one of the spheres vanishes at at the minimal radius
while the other remains of finite size. Furthermore, if we choose to
reduce along an isometry embedded in the vanishing sphere, we need to
recall that the vanishing of the M-theory circle indicates the
presence of D6-branes. Thus the reduction along
$\partial_{\tilde{\phi}}$ yields a deformed conifold with a D6-brane
at $\rho = \rho_0$ extending along the Minkowski directions and wrapping the
non-vanishing $S^3$. If one mods out the $\Lie{U}(1)$ by
$\mathbb{Z}_{N_c}$ before reducing, the corresponding geometry is that
of $N_c$ branes. The other two reductions include
non-singular $\Lie{U}(1)$'s, so we end up with resolved
conifolds. As the M-theory circle is non-singular, there is no
D6-brane. There is $F_2$ flux though on the finite-size two-sphere. The
different geometries are related by a flop transition between the
resolved conifolds and the conifold transition between the deformed
and the resolved ones.

In the context of gauge/string duality, the deformed conifold corresponds to the weak 't Hooft coupling regime, while the resolved one is to be considered for large 't Hooft coupling. Thus the latter provides the appropriate
supergravity dual. M-theory realizes the conifold dualities via the
aforementioned moduli space of solutions. See
\cite{Atiyah:2000zz,Edelstein:2001pu,Atiyah:2001qf}.

\paragraph{Scherk-Schwarz gauge}
In what follows we will study the reduction along $\partial_\psi
+ \partial_{\tilde{\psi}}$. In the context of the flavoring problem of section
\ref{sec:smeared-sources_in_iia} one expects the system to be best
described by one of the resolved conifold geometries with additional
flavor branes. Therefore, out of the three isometries discussed
$\partial_\phi$ and $\partial_\psi + \partial_{\tilde{\psi}}$ are the
obvious choices. We selected the latter as it leads to simpler
equations in type IIA. The choice made here does affect the flavoring
problem, yet not our results on the M-theory lift. As we are
interested in the reduction of  tangent-space quantities, we need to
transform the elfbein to Scherk-Schwarz gauge
\begin{equation}
  \label{eq:Scherk-Schwarz_gauge}
  \begin{aligned}
        \hat{e}_M^A &=  \begin{pmatrix}
                      e^{-\frac{1}{3}\Phi} e_\mu^a &
                      e^{\frac{2}{3}\Phi} A_\mu \\
                      0 & e^{\frac{2}{3}\Phi}
                    \end{pmatrix}_{MA}
    &
    \hat{E}_A^M &=  \begin{pmatrix}
                      e^{\frac{1}{3}\Phi} E_a^\mu &
                      -e^{\frac{1}{3}\Phi} A_a \\
                      0 & e^{-\frac{2}{3}\Phi}
                    \end{pmatrix}_{AM}
  \end{aligned}
\end{equation}
To obtain the gauge (\ref{eq:Scherk-Schwarz_gauge}) from
(\ref{eq:d=11_unflavored_elfbein_(brandhuber)}), we perform the
following gauge transformation:
\begin{equation*}
  \Lambda = \Lambda^{(3)} \Lambda^{(2)} \Lambda^{(1)}
\end{equation*}
with the individual transformations
$\Lambda^{(1)}, \Lambda^{(2)}, \Lambda^{(3)}$ being
\begin{equation}
    \label{eq:Lorentz_transformations}
    \begin{aligned}
      \Lambda^{(1)} &= \left(
        \begin{smallmatrix}
          \mathbb{I}_{9 \times 9} & & \\
          & \cos\alpha & -\sin\alpha \\
          & \sin\alpha & \cos\alpha
        \end{smallmatrix}
      \right) \\
      \Lambda^{(2)} &= \left(
        \begin{smallmatrix}
          \mathbb{I}_{5 \times 5} & & & & & \\
          & \cos\frac{\psi_+}{2} & -\sin\frac{\psi_+}{2} & & & \\
          & \sin\frac{\psi_+}{2} & \cos\frac{\psi_+}{2} & & & \\
          & & & \cos\frac{\psi_+}{2} & -\sin\frac{\psi_+}{2} & \\
          & & & \sin\frac{\psi_+}{2} & \cos\frac{\psi_+}{2} & \\
          & & & & & \mathbb{I}_{2 \times 2}
        \end{smallmatrix}     \right) \\
      \Lambda^{(3)} &= \left(
        \begin{smallmatrix}
          \mathbb{I}_{6 \times 6} & & & & \\
          & \cos\alpha & 0 & \sin\alpha & \\
          & 0 & 1 & 0 & \\
          & -\sin\alpha & 0 & \cos\alpha & \\
          & & & & \mathbb{I}_{2 \times 2}
        \end{smallmatrix}
      \right)
    \end{aligned}
\end{equation}
and all other entries zero. Here we defined
\begin{equation}
  \begin{aligned}
    \cos\alpha(\rho) &= \frac{D (1-g)}{\sqrt{B^2 + (1-g)^2 D^2}} &\qquad
    \psi_+ &= \psi + \tilde{\psi} \\
    \sin\alpha(\rho) &= \frac{B}{\sqrt{B^2 + (1-g)^2 D^2}} &\qquad
    \psi_- &= \psi - \tilde{\psi}
  \end{aligned}
\end{equation}
In principle one needs only $\Lambda^{(1)}$ and $\Lambda^{(2)}$ to obtain
Scherk-Schwarz gauge; yet without $\Lambda^{(3)}$ the new projections
satisfied by the SUSY spinor would be linear
combinations of the old ones
(\ref{eq:m-thy_SUSY-spinor_non-reducible-gauge}) with coefficients
$\cos\alpha, \sin\alpha$. As it is, the form of the SUSY
projections remains invariant under $\Lambda$. I.e.
\begin{equation}
  \label{eq:M-Theory_SUSY-projections_reducible_and_nice}
  \begin{aligned}
    \hat{\Gamma}^{1234} \hat{\epsilon} &= \hat{\epsilon} &
    \hat{\Gamma}^{1356} \hat{\epsilon} &= -\hat{\epsilon} &
    \hat{\Gamma}^{\rho 126} \hat{\epsilon} &= -\hat{\epsilon}
  \end{aligned}
\end{equation}
Thus the $G_2$-structure (\ref{eq:m-thy_G2-structure_no-flavors})
remains formally the same, with the vielbeins $\tilde{e}^A$ now
replaced by $\hat{e}^A$. A disadvantage of the reducible gauge is that
the new vielbein is rather complicated.

\paragraph{Dimensional reduction and type IIA string theory}
The resulting type IIA vielbein is given by
\begin{subequations}  \label{eq:iia-background}
\begin{align}
    e^\mu &= e^{\frac{1}{3}\Phi} \td x^\mu \\
    e^\rho &= e^{\frac{1}{3}\Phi} E \td\rho \\
    e^1 &= e^{\frac{1}{3}\Phi} A (\cos\frac{\psi_-}{2} \td\theta
    + \sin\theta \sin\frac{\psi_-}{2} \td\phi) \\
    e^2 &= e^{\frac{1}{3}\Phi} A \cos\alpha
    (\cos\frac{\psi_-}{2} \sin\theta \td\phi - \sin\frac{\psi_-}{2}
    \td\theta) \notag \\
    &+ e^{\frac{1}{3}\Phi} C \sin\alpha \left\lbrack
      \cos\frac{\psi_-}{2} ( \sin\tilde{\theta} \td\tilde{\phi} - f
      \sin\theta \td\phi) + \sin\frac{\psi_-}{2} (\td\tilde{\theta} +
      f \td\theta) \right\rbrack\\
    e^3 &= e^{\frac{1}{3}\Phi} C \left\lbrack
      \cos\frac{\psi_-}{2} ( \td\tilde{\theta} - f \td\theta ) -
      \sin\frac{\psi_-}{2} (f \sin\theta \td\phi + \sin\tilde{\theta}
      \td\tilde{\phi}) \right\rbrack \\
    e^4 &= -e^{\frac{1}{3}\Phi} A \sin\alpha
    (\cos\frac{\psi_-}{2} \sin\theta \td\phi - \sin\frac{\psi_-}{2}
    \td\theta) \notag \\
    &+ e^{\frac{1}{3}\Phi} C \cos\alpha \left\lbrack
      \cos\frac{\psi_-}{2} ( \sin\tilde{\theta} \td\tilde{\phi} - f
      \sin\theta \td\phi) + \sin\frac{\psi_-}{2} (\td\tilde{\theta} +
      f \td\theta) \right\rbrack \\
    e^5 &= e^{\frac{1}{3}\Phi} D \sin\alpha ( \cos\theta \td\phi
    - \cos\tilde{\theta} \td\tilde{\phi} + \td\psi_-)
\end{align}
\end{subequations}
While the dilaton and gauge potential are
\begin{equation}
\begin{aligned}
    e^{\frac{2}{3}\Phi} &= \frac{B}{2\sin\alpha} =
    \frac{D(1-g)}{2\cos\alpha} \\
    A_{(1)} &= \cos\theta \td\phi + \cos\tilde{\theta} \td\tilde{\phi}
    + \frac{B^2 - D^2(1-g^2)}{B^2+(1-g)^2 D^2} (\cos\theta \td\phi -
    \cos\tilde{\theta} \td\tilde{\phi} + \td\psi_-) \\
    &= \cos\theta \td\phi + \cos\tilde{\theta} \td\tilde{\phi} +
    (\sin^2 \alpha - \frac{1+g}{1-g}\cos^2\alpha) (\cos\theta \td\phi -
    \cos\tilde{\theta} \td\tilde{\phi} + \td\psi_-)
  \end{aligned}
\end{equation}
Using $\hat{\Gamma}^{10} = \Gamma^{11}$, the reduction of the SUSY
projections takes a more pleasing form:
\begin{equation}
  \label{eq:IIA_SUSY-projections}
  \begin{aligned}
    \Gamma^{1234} \epsilon &= \epsilon &
    \Gamma^{135} \Gamma^{11} \epsilon &= -\epsilon &
    \Gamma^{\rho 12} \Gamma^{11} \epsilon &= -\epsilon
  \end{aligned}
\end{equation}
This allows us to calculate the generalized calibration form for
D6-branes in this background.
\begin{equation}
  \label{eq:IIA_calibration_form_in_terms_of_vielbeins}
  \phi_{\text{D6}} = (\bar{\epsilon} \Gamma_{a_0 \dots a_6} \epsilon)
  e^{a_0 \dots a_6} = e^{x^0x^1x^2x^3} \wedge ( e^{125} - e^{345} -
  e^{\rho24} - e^{\rho13} )
\end{equation}
Note that the internal three-form part of this is up to some overall dilaton
factor identical to that part of the $G_2$-structure
(\ref{eq:m-thy_G2-structure_no-flavors}) independent of $\hat{e}^6$.

\paragraph{$G$-structures}
\label{sec:g-structures}

In terms of $G$-structures the situation in type IIA is the
following. Because we preserve four supercharges, we expect space-time to
carry an $\Lie{SU}(3)$-structure. As it was shown in
\cite{Kaste:2002xs}, it can be directly derived from the $G_2$-structure of the KK-lift. Centerpiece of that reduction are the relations
\begin{equation} \label{eq:J_and_Psi_from_11d}
  \begin{aligned}
    J &= (\hat{\phi}_{G_2})_{ab6} e^{ab} \\
    \Psi &= (\hat{\phi}_{G_2})_{abc} e^{abc}
  \end{aligned}
\end{equation}
For the six-dimensional internal manifold, $J$ defines an almost
complex structure, with respect to which we can define from $\Psi$ a
$(3,0)$-form $\Omega$ as
\begin{equation}
  \Omega = \Psi - i *_6 \Psi
\end{equation}
These satisfy the equations
\begin{equation}
  \begin{aligned}
    J \wedge \Omega &= 0 \\
    J \wedge J \wedge J &= \frac{3i}{4} \Omega \wedge \bar{\Omega}
  \end{aligned}
\end{equation}
In the case at hand we have
\begin{equation}
  \begin{aligned}
    J &= e^{\rho5} + e^{14} - e^{23} \\
    \Psi &= e^{\rho13} + e^{\rho 24} + e^{345} - e^{125} 
  \end{aligned}
\end{equation}
which gives
\begin{equation}
  \Omega = \Psi - i *_6 \Psi = (e^{\rho} + i e^5) \wedge (e^1 + i e^4)
  \wedge (e^3 + i e^2)
\end{equation}

Thinking about lifting from ten to eleven dimensions, we can invert
equations \eqref{eq:J_and_Psi_from_11d} to express the eleven-dimensional
$G_2$-structure in terms of the ten-dimensional quantities:
\begin{equation}
  \label{eq:relation_of_G2_and_SU3-structures}
  \begin{aligned}
    \hat{\phi}_{G_2} &= e^{-\Phi} \Psi + e^{-\frac{2}{3}\Phi} J \wedge
    \hat{e}^{6} \\
    *_7 \hat{\phi}_{G_2} &= -\frac{1}{2} e^{-4\Phi/3} J \wedge J +
    e^{-\Phi} (*_6 \Psi) \wedge \hat{e}^{6}
  \end{aligned}
\end{equation}
As previously stated, Ricci flatness, preservation of four supercharges
and absence of four-form flux in eleven dimensions guarantee the
$G_2$-holonomy of the internal manifold. This translates in the
closure and co-closure of $\hat{\phi}_{G_2}$. As the fibration of the
M-theory circle over the ten-dimensional base is non-trivial, one
obtains non-vanishing two-form flux upon reduction to type IIA. Hence
the internal six-dimensional manifold does not have
$\Lie{SU}(3)$-holonomy due to its intrinsic torsion. This means that
the forms $J$ and $\Omega$ are not both closed. The relation they will obey
can be derived from the closure and co-closure of $\hat{\phi}_{G_2}$
thanks to
\eqref{eq:relation_of_G2_and_SU3-structures}
\begin{equation}\label{eq:linking_structure_equations}
  \begin{aligned}
    \td \hat{\phi}_{G_2} &= \td (e^{-\Phi} \Psi) + \td J \wedge
    (A_{(1)} + \td \psi_+) + J \wedge \td A_{(1)} = 0 \\
    \td *_7 \hat{\phi}_{G_2} &= -\frac{1}{2} \td (e^{-4\Phi/3}
    J \wedge J) + \td (e^{-\Phi/3} *_6 \Psi) \wedge (A_{(1)} + \td
    \psi_+) \\ 
    &\quad - e^{-\Phi/3} (*_6 \Psi) \wedge \td A_{(1)} = 0
  \end{aligned}
\end{equation}
We know that none of the type IIA quantities depends on
$\psi_+$. Hence, the contribution to the previous equations coming
from $\td \psi_+$ must cancel by itself. It gives
\begin{equation} \label{eq:IIABPSfrom11d}
	\begin{aligned}
		0 &= \td J \\
		0 &= \td (e^{-\Phi/3} *_6 \Psi) \\
		0 &= \td (e^{-\Phi} \Psi) + J \wedge \td A_{(1)} \\
		0 &= -\frac{1}{2} \td (e^{-4\Phi/3} J \wedge J) -
                e^{-\Phi/3} (*_6 \Psi) \wedge \td A_{(1)}
	\end{aligned}
\end{equation}
These equations can be rephrased (following \cite{Kaste:2002xs} for
example) as
\begin{equation}
  \begin{aligned}\label{eq:IIABPS_geometric_formulation}
    \td J &= 0 \\
    \td \Phi &= \frac{3}{4} e^{\Phi} \td A_{(1)} \lrcorner (*_6 \Psi) \\
    J \lrcorner \td A_{(1)} &= 0
  \end{aligned}
\end{equation}
where
\begin{equation}
  G_{(p)} \lrcorner H_{(p+q)} = \frac{1}{p!} G^{\mu_1 ... \mu_p}
  H_{\mu_1 ... \mu_p \mu_{p+1} ... \mu_{p+q}} \td x^{\mu_{p+1}} \wedge
  ... \wedge \td x^{\mu_{p+q}}
\end{equation}
We described in this section the construction of a type IIA background
from the reduction of eleven-dimensional supergravity. We also derived
the equations imposed on the structure by supersymmetry. Now we turn
to the problem of adding backreacting flavors in this ten-dimensional
context.

\subsection{Smeared sources in type IIA supergravity}
\label{sec:smeared-sources_in_iia}

\subsubsection{The source-modified first-order system}
\label{sec:source-modified_BPSs_in_typeIIA}

Applying the method developed in \cite{Gaillard:2008wt}, we are now
addressing the problem of flavoring the type IIA background obtained
in the previous section. It means that we are looking for a solution
to the following action, describing the backreaction of smeared
D6-brane sources in a type IIA background:
\begin{equation}\label{eq:action_d6-couling_to_f8}
  S = S_{IIA} - T_6 \int \big(e^{-\Phi} \phi_{\text{D6}} - A_{(7)}
  \big) \wedge \Xi_{(3)}
\end{equation}
where $S_{IIA}$ is the type IIA supergravity action,
$\phi_{\text{D6}}$ is the calibration form corresponding to
supersymmetric D6-branes, $A_{(7)}$ is the seven-form potential and
$\Xi_{(3)}$ is the smearing form, representing the distribution of
sources. The sources in \eqref{eq:action_d6-couling_to_f8} modify the
standard type IIA equations of motion and Bianchi identities to
\begin{equation}\label{eq:source_modified_eoms_in_typeIIA}
  \begin{aligned}
    \td F_{(2)} &= -(2 \kappa_{10}^2 T_6) \Xi_{(3)} \\
    0 &= \td *_{10} F_{(2)} \\
    0 &= \frac{1}{\sqrt{-g}} \partial_\kappa ( \sqrt{-g} g^{\kappa
      \lambda} e^{-2\Phi} \partial_\lambda \Phi ) - \frac{3}{8} F^2 -
    \frac{3}{4} e^{-\Phi} \Xi \lrcorner (*_{10} \phi_{D6}) \\
    R_{\mu \nu} &= - 2 \nabla_\mu \partial_\nu \Phi +
    \frac{e^{2\Phi}}{2} (F_{\mu\kappa} F_\nu^{\phantom{\nu}\kappa} -
    \frac{1}{4} g_{\mu\nu} F_{(2)}^2 ) \\
    &\quad + \frac{e^{\Phi}}{4} \big(
    (*_{10} \phi_{D6})_{\mu}^{\phantom{\mu} \kappa \lambda} \Xi_{\nu \kappa
      \lambda} - g_{\mu\nu} \Xi \lrcorner (*_{10} \phi_{D6}) \big)
  \end{aligned}
\end{equation}
Fortunately, the flavoring procedure does not require us to explicitly
solve the complete second-order system. Due to the standard
integrability arguments (\cite{Lust:2004ig,Koerber:2007hd}), it is
sufficient to satisfy the Bianchi identities along with the first-order BPS equations.\footnote{Technically there are further mild
  assumptions to satisfy. I.e.~one needs the $(0,\mu)$ components of the
  Einstein equation to vanish explicitly.} However, in section
\ref{sec:back-11d}
we will show how to derive the second-order system directly from M-theory.

The metric ansatz is given by the vielbein \eqref{eq:iia-background}
and the dilaton is assumed to depend only on the radial coordinate
$\rho$.
The calibration associated with $\kappa$-symmetric D6-branes is given
by \eqref{eq:IIA_calibration_form_in_terms_of_vielbeins} which is
\begin{equation}
	\phi_{\text{D6}} = e^{x^0x^1x^2x^3} \wedge \Psi
\end{equation}
Supersymmetry requires the two-form flux to obey the generalized
calibration condition
\begin{equation}
	\label{eq:generalized_calibration_condition}
	*_{10} \td (e^{-\Phi} \phi_{\text{D6}}) = F_{(2)}
\end{equation}
This tells us that the most general ansatz for $F_{(2)}$  is
\begin{equation}
  \label{eq:F_ansatz}
  F_{(2)} = e^{-4\Phi/3} \big( F_{\rho 5}(\rho) e^{\rho 5} +
  F_{12}(\rho) e^{12} + F_{14}(\rho) e^{14} + F_{23}(\rho) e^{23} +
  F_{34}(\rho) e^{34} \big)
\end{equation}
The conditions given by
supersymmetry on this $\Lie{SU}(3)$ geometry with intrinsic torsion
are still given by (see end of section \ref{sec:g-structures})
\begin{equation}
  \begin{aligned}
    \td J &= 0 \\
    \td \Phi &= \frac{3}{4} e^{\Phi} F_{(2)} \lrcorner (*_6 \Psi) \\
    J \lrcorner F_{(2)} &= 0
  \end{aligned}
\end{equation}
where we have now replaced $\td A_{(1)}$ by $F_{(2)}$, thus allowing
for $\td F_{(2)} \neq 0$, as necessary for D6 sources.
Together with the generalized calibration condition
\eqref{eq:generalized_calibration_condition}, these equations give the first-order equations the system must satisfy:
\begin{equation}
  \label{eq:10D_BPS}
  \begin{aligned}
    f &= \frac{A}{C \tan \alpha} \\
    \alpha' &= \frac{E}{2} \left( \frac{2}{D \tan \alpha} -\frac{D}{C^2 \tan \alpha} + \frac{D \cos \alpha \sin \alpha}{A^2}- 2 F_{23} \right) \\
    A' &= \frac{E}{2} \left( \frac{A}{D \tan^2 \alpha} + \frac{D}{A} - \frac{A D}{C^2 \tan^2 \alpha} - \frac{2 A F_{23}}{\tan \alpha}- A F_{34} \right) \\
    C' &= \frac{E}{2} \left( -\frac{C}{D \tan^2 \alpha} + \frac{D}{C} - C F_{34} \right) \\
    D' &= \frac{E}{2} \left( -\frac{2 D^2}{C^2} -\frac{D^2}{A^2} + \frac{D F_{23}}{\tan \alpha} + 2 \right) \\
    \Phi' &= \frac{3 E}{2} \left( -\frac{D \cos^2 \alpha}{2A^2} + \frac{D}{2C^2 \tan^2 \alpha} + \frac{F_{23}}{\tan \alpha} + F_{34} \right) \\
    F_{\rho 5} &= \frac{D \cos \alpha \sin \alpha}{A^2} - \frac{D}{C^2 \tan \alpha} \\
    F_{12} &= \frac{D \cos^2 \alpha}{A^2} - \frac{D}{C^2 \tan^2 \alpha} - \frac{2 F_{23}}{\tan \alpha} - F_{34} \\
    F_{14} &= \frac{D}{C^2 \tan \alpha} - \frac{D \cos \alpha \sin \alpha}{A^2} + F_{23}
  \end{aligned}
\end{equation}
As mentioned before, the modified equations of motion relate the
smearing form to the two-form flux.
\begin{equation}
	\td F_{(2)} = -2 \kappa^2_{10} T_6 \Xi
\end{equation}
This equation, combined with \eqref{eq:F_ansatz} and
\eqref{eq:10D_BPS}, tells us that the most general ansatz for $\Xi$ is
\begin{equation}
  \Xi = e^{-5\Phi/3} \big( \Xi_1 (\rho) e^{\rho34} + \Xi_2
  (\rho) (e^{\rho23} +  e^{\rho14}) + \Xi_3 (\rho) e^{\rho12} +
  \Xi_4(\rho) (e^{135} + e^{245}) \big)
\end{equation}
with
\begin{equation}
  \begin{aligned}
    \Xi_3 &= - \Xi_1 - \frac{2 \Xi_2}{\tan \alpha} \\
    \Xi_4 &= \frac{F_{34}}{2 \kappa^2_{10} T_6 D \sin^2 \alpha}
  \end{aligned}
\end{equation}
and the additional conditions
\begin{equation}
  \label{eq:F_Derivatives_Conditions}
  \begin{aligned}
    F_{23}' = &E \left( -\frac{F_{34}}{D \tan \alpha} - \frac{D F_{23}
        \cos^2 \alpha + D F_{34} \cos \alpha \sin \alpha}{A^2} -
      \frac{D^2 \cos \alpha \sin \alpha}{A^2 C^2} + \frac{2
        F_{23}^2}{\tan \alpha} \right. \\
    &\left. + \frac{D F_{23} \cos(2\alpha) + D F_{34} \sin \alpha \cos
        \alpha}{C^2 \sin^2 \alpha} + \frac{D^2 \cos \alpha}{C^4 \sin
        \alpha} + 3 F_{34} F_{23} - 2 \kappa^2_{10} T_6 \Xi_2
    \right)\\
    F_{34}' = &E \left( \frac{F_{34}}{D \tan^2 \alpha} - \frac{D
        F_{34}}{2 A^2} + \frac{D F_{34} \cos (2 \alpha)}{2 C^2 \sin^2
        \alpha} + \frac{F_{34} F_{23}}{\tan \alpha} + 2 F_{34}^2 - 2
      \kappa^2_{10} T_6 \Xi_1 \right)
	\end{aligned}
\end{equation}
One can verify explicitly that any solution to equations
(\ref{eq:10D_BPS}) and (\ref{eq:F_Derivatives_Conditions})
automatically verifies the source-modified equations of motion
(\ref{eq:source_modified_eoms_in_typeIIA}).

As we want to interpret the two-form flux $F_{(2)}$ as created by
brane sources, we need the flux to be quantized, obeying $\int_{S^2}
F_{(2)} = 2 \pi N_c$. $S^2$ is a suitable two-cycle surrounding the
branes in the transverse, three-dimensional space. This adds
constraints on $\Xi$ and $F_{(2)}$:
\begin{equation}
  \begin{aligned}
    \Xi_1 &= \Xi_2 \tan \alpha \\
    F_{23} &= \frac{-A^2 D + C^4 F_{34} \sin^2 \alpha + C^2 (2N_c
      e^{\frac{2}{3}\Phi} \sin \alpha \tan \alpha + D \sin^2 \alpha +
      A^2 F_{34})}{(A^2 C^2 + C^4 \sin^2 \alpha) \tan \alpha}
  \end{aligned}
\end{equation}
that are compatible with the equation \eqref{eq:F_Derivatives_Conditions}.

\subsection{Finding a solution}
\label{sec:discussing_the_iia_solution}

In this section, we present an analytic solution to the previous
system of first-order equations. We will notice
that this solution corresponds to the addition of sources in the
singular conifold. First, we can directly solve one of the equations
in \eqref{eq:10D_BPS}:
\begin{equation}
	D = e^{\frac{2}{3} \Phi} \frac{N_c C^2 \sin \alpha \tan \alpha}{A^2}
\end{equation}
Let us now specialize to the case $\Xi_2 = 0$. We see that this reduces the freedom of the smearing form to
\begin{equation}
	\Xi_{(3)} = \frac{e^{-5\Phi/3} F_{34}}{2 \kappa^2_{10} T_6 D
          \sin^2 \alpha} (e^{135} + e^{245})
\end{equation}
The branes smeared with this particular form would correspond to
branes extended in the radial direction $\rho$ in a trivial way. For a
discussion of $\kappa$-symmetric brane embeddings in this geometry,
see appendix \ref{sec:brane_embeddings}. 
This simplification enables us to solve the equation for the last
unknown component of the two-form flux $F_{(2)}$:
\begin{equation}
  F_{34} = e^{\frac{2}{3} \Phi} \frac{N_f \sin \alpha}{A C}
\end{equation}
where $N_f$ is a constant of integration related to the number of
flavors in the dual field theory. We now suppose that the two-form
flux is independent of the radial coordinate $\rho$, a property
verified in other examples of string duals. This imposes that 
\begin{equation}
  A^2 = C^2 \sin^2 \alpha
\end{equation}
Finally, we assume $f$ to be constant. A look at the original metric
\eqref{eq:iia-background} tells us that $f$ parameterizes the fibration
between the two spheres -- this becomes rather more obvious in
\eqref{eq:d=11_unflavored_elfbein_(brandhuber)}. Thus if $f$ is
independent of $\rho$, the fibration does not change if we flow along
the radial direction. Then we can solve the full BPS system
analytically, and we find:
\begin{equation}
	\begin{aligned}
		D^2 &= e^{\frac{4}{3}\Phi} \frac{N_c^2}{f^2} \\
    A^2 &= e^{\frac{4}{3}\Phi} \frac{4 N_c^2 (1-f^2)^2}{3 f^2} \\
    C^2 &= e^{\frac{4}{3}\Phi} \frac{4 N_c^2 (1-f^2)}{3 f^2} \\
    E^2 &= \frac{16 N_c^2 (1-f^2)^2}{f^2} \lbrack (e^{\frac{2}{3}\Phi})^\prime \rbrack^2 \\
		\cos \alpha &= f \\
		N_f &= \pm \frac{N_c (4f^2-1)}{3f}
	\end{aligned}
\end{equation}
where $0<f<1$. 
The two-form flux is
\begin{equation}
	\begin{aligned}
		F_{(2)} = &- N_c \Big( \sin \theta \td \theta \wedge \td \phi + \sin \tilde{\theta} \td \tilde{\theta} \wedge \td \tilde{\phi} \Big) \\
		&+ N_f \sin \psi_- \Big(\td \theta \wedge \td \tilde{\theta} + \sin \theta \sin \tilde{\theta} \td \phi \wedge \td \tilde{\phi} \Big) \\
		&+ N_f \cos \psi_- \Big( \sin \tilde{\theta} \td \theta \wedge \td \tilde{\phi} + \sin \theta \td \tilde{\theta} \wedge \td \phi \Big)
	\end{aligned}
\end{equation}

At this point we notice that we can write the metric explicitly as a
cone upon redefinition of the radial coordinate. We take
\begin{equation}
	r = \frac{4N_c (1-f^2)}{f} e^{2\Phi/3}
\end{equation}
then $\td r^2 = E^2 \td \rho^2$ and the metric is
\begin{equation}
  \td s_{IIA}^2 = e^{2\Phi/3} \Big( \td x^2_{1,3} + \td r^2 +
  r^2 \td \Omega_{int}^2 \Big) 
\end{equation}
where
\begin{equation}
  \begin{aligned}
    \td \Omega_{int}^2 = &\frac{1}{12} (\td\theta^2 + \sin^2\theta \td\phi^2) +\frac{1}{12 (1-f^2)} \lbrack(\omega_1 - f \td\theta)^2 + (\omega_2 - f \sin\theta \td\phi)^2 \rbrack \\
    &+ \frac{1}{16 (1-f^2)} (\omega_3 - \cos\theta \td\phi )^2
  \end{aligned}
\end{equation}
We can first notice that taking the limit $N_f \to 0$ for this solution gives the singular
conifold. It indeed correspond to taking $f \to \frac{1}{2}$, giving
\begin{equation}
  \begin{aligned}
    \td s_{N_f \to 0}^2 = &\frac{r}{6N_c} \Big( \td x^2_{1,3} + \td r^2 + \frac{r^2}{12} (\td\theta^2 + \sin^2\theta \td\phi^2) \\
    & +\frac{r^2}{9} \lbrack(\omega_1 - \frac{1}{2} \td\theta)^2 + (\omega_2 - \frac{1}{2} \sin\theta \td\phi)^2 \rbrack + \frac{r^2}{12} (\omega_3 - \cos\theta \td\phi )^2 \Big)
  \end{aligned}
\end{equation}
Secondly, we have quantization of the two-form color flux, which is
necessary for the gauge/string duality.

The interpretion of
the additional flavor terms to the flux is not clear. A look at the
solution and appendix \ref{sec:brane_embeddings} prompts us to suspect
that the interpretation of the sources as being due to flavor
branes is more straightforward if one reduces along
$\partial_\phi$. It should be interesting to consider the solution at
hand in
the context of conifold transitions though. Of course, this is just
one solution of the BPS equations of this particular dimensional
reduction. Other solutions might also present interesting
properties. In any way, we leave the study and interpretation of
flavored solutions to future work, and turn in the following back to
the problem of the M-theory lift.

\section{Back to M-theory}
\label{sec:back-11d}

Having studied the flavoring problem of D6-branes in the background
(\ref{eq:iia-background}) in the previous section, we have 
sufficient intuition to turn back towards the more general case of
smeared D6 sources in M-theory. The discussion here is fairly generic
and requires only the presence of the various $G$-structures as well as
the overall topology $\mathbb{R}^{1,3} \times \mathcal{M}$.

\subsection{Lifting the SUSY variations}
\label{sec:lifting_the_SUSY_variations}

\subsubsection{\linkFixer{The $G_2$-structure}{The G2-structure}}
\label{sec:the_G2-structure}

Our considerations in the introduction about the loss of Ricci
flatness prompted us to consider the appearance of intrinsic torsion.
So we will begin our attempt at finding a candidate M-theory lift with
magnetic $A_{(1)}$ sources by studying the ten and eleven-dimensional
$G$-structures. Originally we were dealing with a
$G_2$-holonomy manifold in eleven dimensions. Then we reduced it
to an $\Lie{SU} (3)$-structure in ten dimensions, following the
equations \eqref{eq:J_and_Psi_from_11d}. After this we flavored the
theory, which changed the
structure equations in ten dimensions \eqref{eq:IIABPSfrom11d} by
replacing $\td A_{(1)}$ by $F_{(2)}$. However, after adding sources in type IIA
supergravity, we have $F_{(2)} \neq \td A_{(1)}$. So, if we try to
lift back to eleven dimensions, we start from
\begin{equation}
  \begin{aligned}
    0 &= \td J \\
    0 &= \td (e^{-\Phi/3} *_6 \Psi) \\
    0 &= \td (e^{-\Phi} \Psi) + J \wedge F_{(2)} \\
    0 &= -\frac{1}{2} \td (e^{-4\Phi/3} J \wedge J) - e^{-\Phi/3} (*_6
    \Psi) \wedge F_{(2)}
  \end{aligned}
\end{equation}
When we then look at the $G_2$-structure we find, combining
\eqref{eq:linking_structure_equations} with the above,
\begin{equation}\label{eq:BPS_equations_for_G2-structure_with_sources}
  \begin{aligned}
    \td \hat{\phi}_{G_2} 
    &= - J \wedge (F_{(2)} - \td A_{(1)}) \\
    \td *_7 \hat{\phi}_{G_2} 
    &= e^{-\Phi/3} (*_6 \Psi) \wedge (F_{(2)} - \td A_{(1)})
  \end{aligned}
\end{equation}
So sources in type IIA supergravity translate in eleven dimensions
to the loss of $G_2$-holonomy and the appearance of torsion
proportional to $F_{(2)} - \td A_{(1)} = B_{(2)}$.

\subsubsection{The SUSY variations}
\label{sec:susy-variations}

The previous section gave a first confirmation of our suspicion that
geometric torsion should allow us to accomodate for the sources in
M-theory. This suggests that all geometric quantities such as
covariant derivatives and curvature tensors should be replaced by
their torsion-modified relatives. Simplest among these is the
covariant derivative, which makes an explicit appearance in the
eleven-dimensional supergravity variation $\delta_{\hat{\epsilon}}
\hat{\psi}_M = D_M \hat{\epsilon}$, which yields the IIA supergravity
variations upon KK-reduction. In appendix
\ref{sec:kk-reduction_with_torsion} we therefore
study how this equation and its Kaluza-Klein reduction change upon
inclusion of a torsion tensor\footnote{
Of course, once we include the torsion and proceed from $D_M
\hat{\epsilon}$ to $D^{(\tau)}_M \hat{\epsilon}$, it is not certain
whether this defines a SUSY variation of a supergravity theory. What
we do know for certain however -- and will show in the following -- is
that the naive dimensional reduction of the usual
eleven-dimensional SUSY variation does not yield the correct type IIA
one and that (\ref{eq:gravition_variation_with_torsion}) gives a first
order differential on the spinor that does reduce to the correct
equations. With this in mind, we write $\delta_{\hat{\epsilon}}
\hat{\psi}_M = D^{(\tau)}_M \epsilon$.
}
$\hat{\tau}$
\begin{equation}
  \label{eq:gravition_variation_with_torsion}
  \delta_{\hat{\epsilon}} \hat{\psi}_M = \partial_M \hat{\epsilon} +
  \frac{1}{4} \hat{\omega}_{MAB} \hat{\Gamma}^{AB} \hat{\epsilon} +
  \frac{1}{4} \hat{\tau}_{MAB} \Gamma^{AB} \hat{\epsilon} \equiv
  D^{(\tau)}_M \epsilon
\end{equation}
The result is given in
(\ref{eq:susy_spinor_reduction_final_result}) and we proceed by
investigating what constraints we have to impose on $\hat{\tau}_{MAB}$
in order for the lower-dimensional variations to include magnetic
sources.

Now from the form of the dilatino variation (Einstein frame),
\begin{equation}
  \label{eq:generic_dilatino_variation_with_torsion}
  \begin{aligned}
    \delta_\epsilon \lambda &=
    \frac{3}{16} \frac{1}{\sqrt{2}} e^{\frac{3}{4}\Phi} ( \td A_{bc}
    + 2 e^{-\frac{3}{2}\Phi} \hat{\tau}_{zbc} ) \Gamma^{bc}
    \epsilon + \frac{\sqrt{2}}{4} (\partial_b \Phi +  \frac{3}{2}
    e^{-\frac{3}{4}\Phi} \hat{\tau}_{zbz} )\Gamma^b
    \Gamma^{11}\epsilon
  \end{aligned}
\end{equation}
it follows that we have to demand $\hat{\tau}_{z a z} = 0$
and $\hat{\tau}_{z bc} = T_6 \kappa_{10}^2
\frac{e^{\frac{3}{2}\Phi}}{2} B_{bc}$, as
(\ref{eq:generic_dilatino_variation_with_torsion}) then takes the form
\begin{equation}
  \label{eq:dilatino_variation_with_torsion}
  \begin{aligned}
    \delta_\epsilon \lambda &=
    \frac{3}{16} \frac{1}{\sqrt{2}} e^{\frac{3}{4}\Phi} F_{bc}
    \Gamma^{bc} \epsilon + \frac{\sqrt{2}}{4} \partial_b \Phi 
    \Gamma^b \Gamma^{11}\epsilon
  \end{aligned}
\end{equation}
with the two-form now no longer closed,
$F = \td A + T_6 \kappa_{10}^2 B$.

Substituting $\hat{\tau}_{zaz}$ and $\hat{\tau}_{zbc}$ into the
gravitino varition of (\ref{eq:susy_spinor_reduction_final_result}) we
see that if we impose
\begin{equation}
  \label{eq:generic_form_of_torsion}
  \begin{aligned}
    \hat{\tau}_{zaz} &= 0 &
    \hat{\tau}_{zbc} &= \frac{e^{\frac{3}{2}\Phi}}{2} B_{bc} &
    \hat{\tau}_{\mu bc} &= \frac{e^{\frac{3}{2}\Phi}}{2} A_\mu B_{bc} &
    \hat{\tau}_{\mu b z} &= -\frac{e^{\frac{3}{4}\Phi}}{2} B_{\mu b}
  \end{aligned}
\end{equation}
the gravitino variation turns also to the desired form
\begin{equation}
  \label{eq:gravition_variation_with_torsion2}
  \begin{aligned}
    \delta_\epsilon \psi_\mu &= \partial_\mu \epsilon  
    + e_\mu^a \frac{1}{4} \omega_{abc} \Gamma^{bc} \epsilon
    + \frac{1}{64} e^{\frac{3}{4}\Phi} e_\mu^a F_{cd} \left(
      \eta_{ab} \Gamma^{bcd} - 14 \delta_a^c \Gamma^d \right)
    \Gamma^{11} \epsilon    
  \end{aligned}
\end{equation}
Equations \eqref{eq:dilatino_variation_with_torsion} and
\eqref{eq:gravition_variation_with_torsion2} are important results. If
one performs a KK-reduction of the original supergravity variation
without torsion, $\delta_{\hat{\epsilon}} \hat{\psi}_M = D_M
\hat{\epsilon}$, one obtains supergravity variations including $\td
A_{(1)}$, yet not $B_{(2)} = F_{(2)} - \td A_{(1)}$. By adding the
torsion term to the eleven-dimensional supergravity variation, we are
able to directly derive the ten-dimensional variations with $F_{(2)}$
instead of $\td A_{(1)}$. Looking back at
(\ref{eq:gravition_variation_with_torsion}) it is fair to say that the
spin connection $\hat{\omega}_{MAB}$ contains $\td A_{(1)}$, while the
torsion carries the $B_{(2)}$ term necessary to complete
$F_{(2)}$. The right-hand side of
\eqref{eq:gravition_variation_with_torsion} is constituted of two
parts. The first two terms are the ones coming from the lift of the
IIA part, and are exactly the terms present in eleven-dimensional
supergravity. The last term, which is the only one involving the
torsion, corresponds to the lift of the contribution of the sources to
the ten-dimensional supergravity variations. Thus, it seems that,
mimicking what happens in ten dimensions, we are in presence of the
usual eleven-dimensional supergravity plus some sources.

Using the torsion-modified covariant derivative for spinors
(\ref{eq:gravition_variation_with_torsion}) we can
also define such an operator $\nabla^{(\tau)}$ for tensors. The
relevant connection coefficients $\Gamma$ are
\begin{equation}
  \label{eq:definition_of_connection_with_torsion}
  \begin{aligned}
      \Gamma^K_{\phantom{K}LM} &= \LevCi{K}{L}{M} +
      K^K_{\phantom{K}LM} \\
      K_{AMB} &= \hat{\tau}_{MAB}
  \end{aligned}
\end{equation}
where $\LevCi{K}{L}{M}$ is the Levi-Civita connection. 
With the help of $\nabla^{(\tau)}$, we can rewrite
equations~(\ref{eq:BPS_equations_for_G2-structure_with_sources}) as
\begin{equation}
  \label{eq:torsion_modified_covariant_derivative_of_g2-structure}
  \begin{aligned}
    \nabla^{(\tau)}_M \hat{\phi}_{G_2} &= 0 \\
    \nabla^{(\tau)}_M (*_7 \hat{\phi}_{G_2}) &= 0
  \end{aligned}
\end{equation}
One should remember that the original BPS equations could be written
geometrically as $\nabla_M \hat{\phi}_{G_2} = 0$ and $\nabla_M
*_7\hat{\phi}_{G_2} = 0$ yet that these ceased to be valid once we
include the sources in ten dimensions -- as we discussed in section
\ref{sec:the_G2-structure}. Equations
(\ref{eq:torsion_modified_covariant_derivative_of_g2-structure}) show
however that these geometric BPS equations remain formally invariant
once we include torsion.

\subsection{The equations of motion}
\label{sec:equations-motion}

We shall finally turn to the search for equations of motion in
M-theory that reduce to the source-modified second-order equations in
type IIA as given in equation
(\ref{eq:source_modified_eoms_in_typeIIA}). To find these equations,
we actually reverse the integrability argument that allowed us to
consider the first instead of the second-order equations in sections
\ref{sec:source-modified_BPSs_in_typeIIA} and
\ref{sec:discussing_the_iia_solution}.

To get an idea of what we are
about to do, let us briefly digress to the simple case without any
flavors or sources. The Bianchi identities are the usual ones, the equation of
motion is simple Ricci flatness, $\hat{R}_{MN} = 0$, and the $G_2$-structure is closed and co-closed. Thus the latter satisfies 
$\nabla_M \hat{\phi}_{G_2} = 0$. Taking the commutator
\begin{equation}\label{eq:commutator_of_g2_derivatives}
  \begin{aligned}
    0 &= \lbrack \nabla_K , \nabla_L \rbrack \hat{\phi}_{G_2 MNP} \\
    &= -\hat{R}^S_{\phantom{S}MKL} \hat{\phi}_{G_2 SNP} -
    \hat{R}^{S}_{\phantom{S}NKL} \hat{\phi}_{G_2 MSP} -
    \hat{R}^S_{\phantom{S}PKL} \hat{\phi}_{G_2 MNS}
  \end{aligned}
\end{equation}
Upon contraction of (\ref{eq:commutator_of_g2_derivatives}) with
$\hat{\phi}_{G_2}$, we
find\footnote{\label{fn:the_footnote_with_the_g2_identities} 
As one can verify by direct calculation using
(\ref{eq:m-thy_G2-structure_no-flavors}), the $G_2$-structure
satisfies
\begin{equation*}
  \label{eq:g2-structure_identity}
  \begin{aligned}
    \hat{\phi}_{G_2 lmn} \hat{\phi}_{G_2}^{\phantom{G_2} kmn} &= 6
    \delta_l^k \\
    \hat{\phi}_{G_2}^{\phantom{G_2}klp} \hat{\phi}_{G_2 mnp} &= 
    (*_7 \hat{\phi}_{G_2})_{mn}^{\phantom{mn}kl} + \delta^k_m \delta^l_n
    - \delta^k_n \delta^l_m
  \end{aligned}
\end{equation*}
$k,l,m,n,p$ denote indices of the seven-dimensional internal manifold.
}
\begin{equation}\label{eq:second_order_equation_satisfied_by_ricci_tensor}
  \begin{aligned}
    0 &= 2 \hat{R}_{KL} + \hat{R}_{MNPL} (*_7
    \hat{\phi}_{G_2})_K^{\phantom{K}MNP}
  \end{aligned}
\end{equation}
In the absence of torsion, $\hat{R}_{MNPL} (*_7
\hat{\phi}_{G_2})_K^{\phantom{K}MNP} = 0$, due to the well-known
symmetries satisfied by the Riemann tensor, 
\begin{equation}\label{eq:standard_symmetries_of_torsion_free_ricci_tensor}
  \begin{aligned}
    \hat{R}_{K \lbrack LMN \rbrack} &= 0 \\
    \hat{R}_{KLMN} &= \hat{R}_{MNKL} = -\hat{R}_{MNLK}
  \end{aligned}
\end{equation}
Therefore, our space-time is Ricci flat and the equations of motion
are satisfied.

After this brief digression, we return to the original problem. Our
aim is to find a suitable equation of motion in M-theory, that reduces
to (\ref{eq:source_modified_eoms_in_typeIIA}) upon dimensional
reduction. For consistency this equation of motion needs to reduce to
simple Ricci flatness in the limit where the type IIA source density
$\Xi$ -- equivalently the torsion $\hat{\tau}$ in M-theory --
vanishes. In opposite to our considerations in the previous paragraph,
the $G_2$-structure does no longer satisfy $\nabla_M \hat{\phi}_{G_2}
= 0$, but instead satisfies $\nabla^{(\tau)}_M \hat{\phi}_{G_2} = 0$. So
we can once more consider the commutator of covariant derivatives. The
identities of footnote \ref{fn:the_footnote_with_the_g2_identities}
used to derive
\eqref{eq:second_order_equation_satisfied_by_ricci_tensor}
still hold, yet
(\ref{eq:standard_symmetries_of_torsion_free_ricci_tensor}) do not,
and we arrive at the main result of this paper, the M-theory lift of
the source-modified equations of motion
\begin{equation}
  \label{eq:second_order_equation_satisfied_by_ricci_tensor_with_torsion}
  \begin{aligned}
    0 &= 2 \hat{R}^{(\tau)}_{KL} + \hat{R}^{(\tau)}_{MNPL} (*_7
    \hat{\phi}_{G_2})_K^{\phantom{K}MNP}
  \end{aligned}
\end{equation}
where $\hat{R}^{(\tau)}$ is the Riemann (Ricci) tensor in the presence
of torsion.

As we have pointed out before, the BPS equations in their geometric
form -- $\nabla^{(\tau)}_M \hat{\phi}_{G_2} = 0$ -- are equivalent to
those obtained from the SUSY spinor $\epsilon$, $D^{(\tau)}_M \epsilon
= 0$. Therefore we could have derived
(\ref{eq:second_order_equation_satisfied_by_ricci_tensor_with_torsion})
also using (\ref{eq:gravition_variation_with_torsion}). A commutator
of covariant derivatives acting on the SUSY spinor yields
\begin{equation}
  \begin{aligned}
    0 &= \hat{R}^{(\tau)}_{CDML} \hat{\Gamma}^{CD} \hat{\epsilon}
  \end{aligned}
\end{equation}
We then contract with $\bar{\hat{\epsilon}}
\hat{\Gamma}_K^{\phantom{K}M}$ and make use of the identity
\begin{equation}
  \label{eq:4-Gamma-identity}
  \begin{aligned}
    \Gamma^A \Gamma^B \Gamma^C \Gamma^D &= \Gamma^{ABCD} + \eta^{AB}
    \Gamma^{CD} - \eta^{CB} \Gamma^{DA} + \eta^{CD} \Gamma^{AB} +
    \eta^{DA} \Gamma^{BC} \\
    &- \eta^{AC} \Gamma^{BC} - \eta^{BD} \Gamma^{AC} + \eta^{AB}
    \eta^{CD} - \eta^{AC} \eta^{BD} + \eta^{AD} \eta^{BC}
  \end{aligned}
\end{equation}
it follows that
\begin{equation}\label{eq:11-dim_eom_derived_from_spinors}
  \begin{aligned}
    0 &= 2 (\bar{\hat{\epsilon}} \hat{\epsilon}) \hat{R}^{(\tau)}_{KL} +
    (\bar{\hat{\epsilon}} \hat{\Gamma}_{K}^{\phantom{K}MNP}
      \hat{\epsilon}) \hat{R}_{MNPL} + \order{\bar{\hat{\epsilon}}
        \hat{\Gamma}^{AB} \hat{\epsilon}}
  \end{aligned}
\end{equation}
The assumptions made about the SUSY spinor $\hat{\epsilon}$ imply that
there is a $G_2$ structure that can be expressed as
\begin{equation}
  *_7\hat{\phi}_{G_2} = (\bar{\hat{\epsilon}} \Gamma_{ABCD} \hat{\epsilon})
  \hat{e}^{ABCD}
\end{equation}
They also imply that all terms of the form
$\bar{\hat{\epsilon}} \hat{\Gamma}^{AB} \hat{\epsilon}$ vanish. Hence
(\ref{eq:second_order_equation_satisfied_by_ricci_tensor_with_torsion})
follows from (\ref{eq:11-dim_eom_derived_from_spinors}).

The equations of motion
\eqref{eq:second_order_equation_satisfied_by_ricci_tensor_with_torsion}
can be rewritten in a more typical and enlightning fashion using the
Einstein tensor
\begin{equation} \label{eq:einstein_equation_with_energy_momentum_tensor}
	\hat{R}_{KL} - \frac{1}{2} \hat{g}_{KL} \hat{R} = \hat{T}_{KL}
\end{equation}
where $\hat{T}_{KL}$ is the energy-momentum tensor of the sources. It
can be written in terms of the torsion as
\begin{equation}
  \begin{aligned}
    &\hat{T}_{KL} = \nabla_L K^M_{\phantom{M} MK} - \nabla_M
    K^M_{\phantom{M} LK} + K^M_{\phantom{M} LP} K^P_{\phantom{P} MK} -
    K^M_{\phantom{M} MP} K^P_{\phantom{P} LK} \\
    &\;\;\;+\frac{1}{2} \big( \nabla_L K_{MPN} - \nabla_P K_{MLN} +
    K_{MLQ} K^Q_{\phantom{Q} PN} - K_{MPQ} K^Q_{\phantom{Q} LN} \big)
    (*_7 \hat{\phi}_{G_2})_K^{\phantom{K}MNP} \\
    &\;\;\;+ \frac{1}{2} \hat{g}_{KL} \big( \nabla_M
    K^{M\phantom{Q}Q}_{\phantom{M}Q} - \nabla_Q
    K^{M\phantom{Q}Q}_{\phantom{M}M} + K^M_{\phantom{M} MP}
    K^{P\phantom{Q}Q}_{\phantom{P}Q} - K^M_{\phantom{M} QP}
    K^{P\phantom{M}Q}_{\phantom{M}M} \big) \\
    &\;\;\;+\frac{1}{2} \hat{g}_{KL} \big( \nabla_P K_{MQN} + K_{MPR}
    K^R_{\phantom{R} QN} \big) (*_7 \hat{\phi}_{G_2})^{QMNP}
  \end{aligned}
\end{equation}
where $K_{MNP}$ is the contorsion tensor (see
\eqref{eq:definition_of_connection_with_torsion}).
From \eqref{eq:einstein_equation_with_energy_momentum_tensor}, we can
see that the Einstein equation we are proposing contains two terms: on
the left-hand side, one has the Einstein tensor one would get from
varying the eleven-dimensional supergravity action with no four-form
flux; on the right-hand side, one has an energy-momentum tensor that
vanishes when the torsion is set to zero. When the torsion vanishes,
so does $\hat{T}$ and one recovers the M-theory Einstein
equation. Writing the equation in this form makes very clear the fact
that the lift of type IIA supergravity with sources is
eleven-dimensional supergravity supplemented by some
sources. Unfortunately, we were not able to find an action that would
be responsible for this energy-momentum tensor. To summarize, we claim
that having sources in ten dimensions corresponds to having an
energy-momentum tensor in eleven dimensions, of the form presented
above.

To verify our claim, we will now
perform the explicit dimensional reduction of
\eqref{eq:second_order_equation_satisfied_by_ricci_tensor_with_torsion},
and show that we recover all the equations of motion of type IIA with
sources. The calculations are -- as so often in supergravity --
straightforward yet tedious. We found \cite{Peeters:2007wn} quite
helpful, yet not essential. The reader not interested in mathematical
details might want to skip ahead to the end of this section, where we
summarize our findings. Notice that in the following, despite the fact
that we dropped the superscript $(\tau)$ for simplicity of notation,
all hatted Riemann and Ricci tensor are considered in the presence of
torsion.

Let us start with the $zz$-component of
(\ref{eq:second_order_equation_satisfied_by_ricci_tensor}). We find
\begin{equation}
  \begin{aligned}
    \hat{R}_{zz} &= -\frac{2}{3} e^{4 \Phi}
    \frac{1}{\sqrt{-g}} \partial_{\mu} (\sqrt{-g} e^{-2
      \Phi} \partial^{\mu} \Phi) + \frac{1}{4} e^{4\Phi} F^2 \\
    (*_7 \hat{\phi}_{G_2})^{\phantom{z} SPK}_z \hat{R}_{SPKz} &= e^{3 \Phi} (*_6 \Psi)
    \lrcorner \td B
  \end{aligned}
\end{equation}
from which it follows that
\begin{equation}
  \begin{aligned}
    0&= 2 \hat{R}_{zz} + (*_7 \hat{\phi}_{G_2})^{\phantom{z} SPK}_z \hat{R}_{SPKz} \\
    &= -\frac{4}{3} e^{4 \Phi} \frac{1}{\sqrt{-g}} \partial_{\mu}
    (\sqrt{-g} e^{-2 \Phi} \partial^{\mu} \Phi) + \frac{1}{2}
    e^{4\Phi} F^2 + e^{3 \Phi} (*_6 \Psi) \lrcorner \td B \\ 
    &= \frac{1}{\sqrt{-g}} \partial_{\mu} (\sqrt{-g} e^{-2
      \Phi} \partial^{\mu} \Phi) - \frac{3}{8} F^2 - \frac{3}{4} e^{-
      \Phi} (*_{10} \phi_{D6}) \lrcorner \Xi
  \end{aligned}
\end{equation}
Here we used that $*_6 \Psi = - *_{10} \phi_{D6}$ and $\td B = \td F = -
\Xi$. And we notice that we find the source-modified ten-dimensional
equation of motion for the dilaton as in
(\ref{eq:source_modified_eoms_in_typeIIA}).

Now we investigate the $\mu z$-component of 
(\ref{eq:second_order_equation_satisfied_by_ricci_tensor}). We find
\begin{equation}
  \begin{aligned}
    \hat{R}_{\mu z} &= -\frac{1}{2} e^{2\Phi} \nabla^{\nu} F_{\nu \mu}
    + A_{\mu} \hat{R}_{zz} \\ 
    (*_7 \hat{\phi}_{G_2})^{\phantom{\mu} SPK}_{\mu} \hat{R}_{SPKz} &= -\frac{1}{6}
    e_{a \mu} (*_7 \hat{\phi}_{G_2})^{abcd} (\td B)_{bcd} + A_{\mu} (e^{3 \Phi}
    (*_6 \Psi) \lrcorner \td B)
  \end{aligned}
\end{equation}
Now we have
\begin{equation}
  \begin{aligned}
    \frac{1}{6} e_{a \mu} (*_7 \hat{\phi}_{G_2})^{abcd} (\td B)_{bcd} &= \frac{1}{6}
    (*_6 J)^{\mu bcd} (\td B)_{bcd} \\
    &= \frac{1}{12} \sqrt{-g_{(6)}} \epsilon^{\alpha \beta \mu \nu
      \rho \sigma} J_{\alpha \beta} (\td B)_{\nu \rho \sigma} \\
    &= \frac{1}{12} \frac{1}{6!} *_6 (\td x^{\mu} \wedge J \wedge \td B) \\
    &= \frac{1}{12} \frac{1}{6!} *_6 [\td x^{\mu} \wedge \td(J \wedge B)] \\
    &= 0
  \end{aligned}
\end{equation}
because supersymmetry tells us that $\td J = 0$ and $\td (J \wedge B)
= \td ( \td \hat{\phi}_{G_2}) = 0$. Thus
\begin{equation}
  \begin{aligned}
    0&= 2 \hat{R}_{\mu z} + (*_7 \hat{\phi}_{G_2})^{\phantom{\mu} SPK}_\mu \hat{R}_{SPKz} \\
    &= - e^{2\Phi} \nabla^{\nu} F_{\nu \mu} + 2 A_{\mu} \hat{R}_{zz}
    -\frac{1}{6} e_{a \mu} (*_7 \hat{\phi}_{G_2})^{abcd} (\td B)_{bcd} + A_{\mu}
    (e^{3 \Phi} (*_6 \Psi) \lrcorner \td B) \\
    &= - e^{2\Phi} \nabla^{\nu} F_{\nu \mu} + A_{\mu} [2 \hat{R}_{zz}
    + e^{3 \Phi} (*_6 \Psi) \lrcorner \td B]
  \end{aligned}
\end{equation}
The term in square brackets is equal to the $zz$-component of
(\ref{eq:second_order_equation_satisfied_by_ricci_tensor}) and the
remaining part corresponds to the Maxwell equation for $F_{(2)}$. 

The $z \nu$-component\footnote{One might suspect this to be identical
  to the $\mu z$-component. Due to the presence of torsion however,
  the Ricci tensor is no longer symmetric and one has to check this
  independently. Interestingly, the Kaluza-Klein reductions of $\mu z$
  and $z \nu$ are already different in the torsion-free case. Here the
  two differ by $F - \td A$ however, which vanishes in source
  and torsion-free geometries.} of
(\ref{eq:second_order_equation_satisfied_by_ricci_tensor}) gives
\begin{equation}
  \begin{aligned}
    \hat{R}_{z \nu} &= -\frac{1}{2} e^{2\Phi} \nabla^{\rho} F_{\rho
      \nu} + A_{\nu} \hat{R}_{zz} + \frac{2}{3} e^{2\Phi} (\td A -
    F)_{\nu \rho} \partial^{\rho} \Phi \\
    (*_7 \hat{\phi}_{G_2})^{\phantom{z} SPK}_{z} \hat{R}_{SPK \nu} &= e^{3\Phi}
    B_{\nu \beta} [F \lrcorner (*_6 \Psi)]^{\beta}  + A_{\nu} (e^{3 \Phi}
    (*_6 \Psi) \lrcorner \td B) \\
    &= \frac{4}{3} e^{2\Phi} B_{\nu \beta} \partial^{\beta} \Phi +
    A_{\nu} (e^{3 \Phi} (*_6 \Psi) \lrcorner \td B)
  \end{aligned}
\end{equation}
with $\td \Phi = \frac{3}{4} e^{\Phi} F \lrcorner (*_6 \Psi)$ due to
sypersymmetry. Putting things together
\begin{equation}
  \begin{aligned}
    0&= 2 \hat{R}_{z \nu} + (*_7 \hat{\phi}_{G_2})^{\phantom{z} SPK}_z \hat{R}_{SPK \nu} \\
    &= - e^{2\Phi} \nabla^{\rho} F_{\rho \nu} + A_{\nu} [2\hat{R}_{zz}
    + e^{3 \Phi} (*_6 \Psi) \lrcorner \td B]
  \end{aligned}
\end{equation}
This agrees with the $\mu z$-component. Let us finally look at
the $\mu \nu$-compo\-nent of
(\ref{eq:second_order_equation_satisfied_by_ricci_tensor}). We have
\begin{equation}
  \begin{aligned}
		\hat{R}_{\mu \nu} &= R_{\mu \nu} + 2 \nabla_{\mu} \partial_{\nu} \Phi - \frac{e^{2 \Phi}}{2} (F_{\mu \rho} (\td A)_{\nu}^{\phantom{\nu} \rho} - \frac{1}{4} g_{\mu \nu} F^2) - \frac{1}{2} A_{\nu} \nabla^{\rho} F_{\rho \mu} + A_{\mu} \hat{R}_{z \nu} \\
		&\quad - \frac{e^{-2\Phi}}{2} g_{\mu \nu} \hat{R}_{zz}
	\end{aligned}
\end{equation}
and
\begin{equation}
  \begin{aligned}
    (*_7 \hat{\phi}_{G_2})^{\phantom{\mu} SPK}_{\mu} \hat{R}_{SPK \nu} &= A_{\mu} [(*_7 \hat{\phi}_{G_2})^{\phantom{z} SPK}_{z} \hat{R}_{SPK \nu}] + \frac{4}{3} e^{\Phi} (*_6 \Psi)_{\mu}^{\phantom{\mu} cd} B_{\nu d} \partial_c \Phi \\
    &- \frac{1}{6} A_{\nu} e^{2\Phi} e_{a \mu} (*_7 \hat{\phi}_{G_2})^{abcd} (\td B)_{bcd} - e^{\Phi} (*_6 \Psi)_{\mu}^{\phantom{\mu} cd} \nabla_d B_{\nu c} \\
    &+ \frac{1}{2} e^{2\Phi} e_{a \mu} (*_7 \hat{\phi}_{G_2})^{abcd} B_{\nu d} F_{cb} - \frac{1}{2} e^{\Phi} (*_6 \Psi)_{\mu}^{\phantom{\mu} cd} \nabla_{\nu} B_{cd}
  \end{aligned}
\end{equation}
Let us first notice that
\begin{equation}
  (*_6 \Psi)_{\mu}^{\phantom{\mu} cd} (\nabla_d B_{\nu c} + \frac{1}{2} \nabla_{\nu} B_{cd}) = \frac{1}{2} (*_6 \Psi)_{\mu}^{\phantom{\mu} cd} (\td B)_{\nu cd}
\end{equation}
Then from previous computation we know that
\begin{equation}
  e_{a \mu} (*_7 \hat{\phi}_{G_2})^{abcd} (\td B)_{bcd} = 0
\end{equation}
Here are formulae that are going to be useful in the following calculations:
\begin{equation}
  \begin{aligned}
    (*_6 \Psi)^f_{\phantom{f}ab} (*_6 \Psi)^{\phantom{cd}f}_{cd} &= \eta_{ac} \eta_{bd} - \eta_{ad} \eta_{bc} + J_{ac} J_{db} + J_{ad} J_{bc} \\
    \frac{1}{2} (J \wedge J)_{abcd} &= 	J_{ab} J_{cd} + J_{ac} J_{db} + J_{ad} J_{bc}
  \end{aligned}
\end{equation}
and once again
\begin{equation}
  \partial_a \Phi = \frac{3}{4} e^{\Phi} (F \lrcorner (*_6 \Psi))_a = \frac{3}{8} e^{\Phi} F^{bc} (*_6 \Psi)_{bca}
\end{equation}
So
\begin{equation}
  \begin{aligned}
    (*_7 \hat{\phi}_{G_2})^{abcd} F_{cb} &= -\frac{1}{2} (J \wedge J)^{abcd} F_{cb} \\
    &= F_{cb} (J^{ab} J^{cd} + J^{ac} J^{db} + J^{ad} J^{bc}) \\
    &= 2 J^{ab} F_{bc} J^{cd}
  \end{aligned}
\end{equation}
because supersymmetry dictates that $F \lrcorner J = 0$. And
\begin{equation}
  \begin{aligned}
    (*_6 \Psi)_{\mu}^{\phantom{\mu} cd} \partial_c \Phi &= -\frac{3}{8} e^{\Phi} F^{fg} (*_6 \Psi)_{\mu}^{\phantom{\mu} dc} (*_6 \Psi)_{fgc} \\
    &= -\frac{3}{4} e^{\Phi} (F_{\mu}^{\phantom{\mu} d} + J_{\mu}^{\phantom{\mu} f} F_{fg} J^{gd})
  \end{aligned}
\end{equation}
So if we now put everything together, we get
\begin{equation}
  \begin{aligned}
    (*_7 \hat{\phi}_{G_2})^{\phantom{\mu} SPK}_{\mu} \hat{R}_{SPK \nu} &= A_{\mu} [(*_7 \hat{\phi}_{G_2})^{\phantom{z} SPK}_{z} \hat{R}_{SPK \nu}] + e^{2\Phi} e_{a \mu} B_{\nu d} J^{ab} F_{bc} J^{cd}  \\
    &- \frac{1}{2} e^{\Phi} (*_6 \Psi)_{\mu}^{\phantom{\mu} cd} (\td B)_{\nu cd} - e^{2\Phi} B_{\nu d} (F_{\mu}^{\phantom{\mu} d} + J_{\mu}^{\phantom{\mu} f} F_{fg} J^{gd}) \\
    &= A_{\mu} [(*_7 \hat{\phi}_{G_2})^{\phantom{z} SPK}_{z} \hat{R}_{SPK \nu}] - \frac{1}{2} e^{\Phi} (*_6 \Psi)_{\mu}^{\phantom{\mu} cd} (\td B)_{\nu cd} \\
    &\quad - e^{2\Phi} F_{\mu}^{\phantom{\mu} d} B_{\nu d}
  \end{aligned}
\end{equation}
So looking finally at the whole picture
\begin{equation}
  \begin{aligned}
    0 &= 2 \hat{R}_{\mu \nu} + (*_7 \hat{\phi}_{G_2})^{\phantom{\mu} SPK}_{\mu} \hat{R}_{SPK \nu} \\
    &= 2 R_{\mu \nu} + 4 \nabla_{\mu} \partial_{\nu} \Phi - e^{2 \Phi} (F_{\mu \rho} (\td A)_{\nu}^{\phantom{\nu} \rho} - \frac{1}{4} g_{\mu \nu} F^2) - A_{\nu} \nabla^{\rho} F_{\rho \mu} - e^{2\Phi} F_{\mu}^{\phantom{\mu} d} B_{\nu d} \\
    &- e^{-2\Phi} g_{\mu \nu} \hat{R}_{zz} + A_{\mu} [(*_7 \hat{\phi}_{G_2})^{\phantom{z} SPK}_{z} \hat{R}_{SPK \nu}] - \frac{1}{2} e^{\Phi} (*_6 \Psi)_{\mu}^{\phantom{\mu} cd} (\td B)_{\nu cd} + A_{\mu} 2 \hat{R}_{z \nu} \\
    &= 2 R_{\mu \nu} + 4 \nabla_{\mu} \partial_{\nu} \Phi - e^{2 \Phi} (F_{\mu \rho} (\td A + B)_{\nu}^{\phantom{\nu} \rho} - \frac{1}{4} g_{\mu \nu} F^2) - A_{\nu} \nabla^{\rho} F_{\rho \mu} \\
    &+ A_{\mu} [2 \hat{R}_{z \nu} + (*_7 \hat{\phi}_{G_2})^{\phantom{z} SPK}_{z} \hat{R}_{SPK \nu}] - e^{-2\Phi} g_{\mu \nu} [\hat{R}_{zz} + \frac{1}{2} e^{3 \Phi} (*_6 \Psi) \lrcorner \td B] \\
    &- \frac{1}{2} e^{\Phi} (*_{10} \phi_{D6})_{\mu}^{\phantom{\mu} \rho \sigma} \Xi_{\nu \rho \sigma} + \frac{1}{2} e^{-2\Phi} g_{\mu \nu} e^{3 \Phi} (*_6 \Psi) \lrcorner \td B
	\end{aligned}
\end{equation}
which gives
\begin{equation}
  \begin{aligned}
   0 &= 2 R_{\mu \nu} + 4 \nabla_{\mu} \partial_{\nu} \Phi - e^{2 \Phi} (F_{\mu \rho} F_{\nu}^{\phantom{\nu} \rho} - \frac{1}{4} g_{\mu \nu} F^2) \\
    &\quad - \frac{1}{2} e^{\Phi} \big((*_{10} \phi_{D6})_{\mu}^{\phantom{\mu} \rho \sigma} \Xi_{\nu \rho \sigma} - g_{\mu \nu} (*_{10} \phi_{D6}) \lrcorner \Xi \big) \\
    &- A_{\nu} \nabla^{\rho} F_{\rho \mu} + A_{\mu} [2 \hat{R}_{z \nu} + (*_7 \hat{\phi}_{G_2})^{\phantom{z} SPK}_{z} \hat{R}_{SPK \nu}] \\
    &\quad - \frac{e^{-2\Phi}}{2} g_{\mu \nu} [2 \hat{R}_{zz} + e^{3 \Phi} (*_{10} \phi_{D6}) \lrcorner \Xi] \\
  \end{aligned}
\end{equation}
where we recognize the first two lines of this equation as being the
Einstein equation of type IIA supergravity with sources and the rest vanishes thanks to other components of
(\ref{eq:second_order_equation_satisfied_by_ricci_tensor}). This
completes the reduction of eleven-dimensional Einstein equations to
the type IIA supergravity equations of motion.

To summarize, in this section we showed that the equation of motion of eleven-dimensional supergravity with torsion \eqref{eq:second_order_equation_satisfied_by_ricci_tensor_with_torsion}, which is given to us by integrability, reduces to the source-modified type IIA supergravity equations of motion \eqref{eq:source_modified_eoms_in_typeIIA}. It thus shows that adding torsion to eleven-dimensional supergravity reduces to adding smeared D6 sources in type IIA supergravity.

\section{Conclusions, future work}
\label{sec:conclusions_and_future_work}

In this paper we have been interested in two related issues: the
addition of D6-branes as smeared sources to a type IIA background, and the lifting of such a system to eleven-dimensional supergravity. We
considered these in the context of 1/8 BPS solutions of the form
$\mathbb{R}^{1,3} \times \mathcal{M}$, a fact represented by the
presence of a $G_2$ or $\Lie{SU}(3)$-structure.

Concerning the problem of the M-theory lift, we showed that ordinary
eleven dimensional-supergravity cannot accomodate for the presence of
the additional sources and argued that a possible solution might lie
in the inclusion of geometric torsion. While our argument was founded
on the observed loss of Ricci flatness in the higher-dimensional
theory, we were able to show by explicit calculation that the
supersymmetry variations take the required form upon addition of
torsion. Moreover, the torsion must take the form
\eqref{eq:generic_form_of_torsion}, related to the distribution
$\Xi_{(3)}$ 
of the sources in the reduced theory. Subsequently we derived a set of
second order equations that could be the equations of motion of some
eleven-dimensional supergravity with torsion, and proved that they
reduce to the type IIA equations of motion with smeared D6-branes. As
we pointed out, this 
work is not in contradiction with the uniqueness of supergravity in eleven
dimensions, because we are only considering a theory that preserves
four supercharges. We did not of course show that there is a well
defined theory in eleven dimensions that is supersymmetric and has the
field content of both eleven dimensional supergravity as well as of
the additional torsion. One should not forget however, that we are not
studying the uplift of $S_{\text{IIA}}$, which is well known, but of
\begin{equation}
  \begin{aligned}
    S &= S_{\text{IIA}} + S_{\text{D$6$-sources}}
  \end{aligned}
\end{equation}
The problem was first addressed in \cite{Bergshoeff:1997gy} whose
authors found a seven-dimensional gauged sigma model action that
reduces to the DBI-term of the D$6$ brane. They were unable to find a
suitable uplift of the Wess-Zumino term however. While this paper does not solve the problem in the sense of
\cite{Bergshoeff:1997gy}, it does succeed in lifting the
ten-dimensional equations of motion to pure eleven-dimensional
geometry. The question whether the results are just an accidental
rewriting of type IIA dynamics in higher-dimensional notation or do
actually point to a higher dimensional supersymmetric theory that
includes torsion is left for further work.

While there is a long history of the uses of torsion in the context of
string theory, the torsion used in papers such as \cite{Strominger:1986uh}
and \cite{Gauntlett:2003cy} is related to the presence of fluxes, not
of sources. Therefore the addition of further torsion is
a rather unorthodox concept. So it is necessary to wonder if we would
not have been able to solve the problem at hand with simpler
methods. As mentioned before, our
argument was based on the loss of Ricci flatness in eleven
dimensions. One might guess that it is possible to use the four-form in
M-theory, $\hat{F}_{(4)}$, to obtain a suitable energy-momentum tensor
to supplement the Einstein equations. This however leads to four and
three-form flux in type IIA, in contradiction with our results
of section \ref{sec:flavoring-iia}. Another possibility would
be to use the KK-monopole action of \cite{Bergshoeff:1997gy}. There the
authors constructed a gauged sigma model action
(\ref{eq:kk-monopole-gauged-sigma-model}) that is the
dimensional uplift of the DBI term of a D6-brane. Using this, one
could try to lift the
action~(\ref{eq:schematic_flavor-modified_action}) to
M-theory. Yet considered in connection with the standard Kaluza-Klein
mechanism, (\ref{eq:schematic_flavor-modified_action}) is an action in
terms of $\td A_{(1)}$, not $F_{(2)}$. So even if one were able to
lift the brane contribution to
(\ref{eq:schematic_flavor-modified_action}), the supergravity part would still be lacking the source contribution. Still, it might be interesting to try to
match the sigma model action \cite{Bergshoeff:1997gy} with the
inclusion of torsion.

As we mentioned before, our calculations here depend on several
assumptions. Most notable among these are the presence of the $G_2$-structure and the Minkowski directions in the metric. Relaxing
both of these would be a very interesting avenue to follow in the future.

One should also be able to extend the considerations of this paper to
the case of KK theories with non-abelian field strengths. While this
is not directly related to the issue of smeared D6-branes, one should
be facing the same difficulties we did; after all the existence of a gauge
potential is again implied by the KK-ansatz.

A further extension and application of the results of this paper lies
outside of string theory. A close look at the considerations made in
section \ref{sec:back-11d} shows that we hardly make any use of string
or M-theory. The setup is merely that of a $\Lie{U}(1)$ Kaluza-Klein
theory in $d$ and $d+1$ dimensions with monopole condensation in the
lower dimensional theory. Hence the results of this paper may be
reexpressed as follows: a monople condensate in a $d$-dimensional
Kaluza-Klein theory might be described as torsion in $d+1$ dimensions.

The other problem studied in this paper is the construction of a gravity dual to $\mathcal{N}=1$, $\Lie{SU}(N_c)$ super Yang-Mills with
flavors. We addressed this in section \ref{sec:flavoring-iia}. Here we
found a system of first-order BPS
equations that describes the addition of D6 sources to the type IIA
background (\ref{eq:iia-background}). At the end of section
\ref{sec:flavoring-iia}, we presented a family of exact solutions. The
detailed study of these, especially concerning the physics of
their gauge theory dual, has not been made and could be of interest as
future work, as well as finding other solutions.

Discussing the addition of flavors to (\ref{eq:iia-background}), we
ignored the complicated issue of conifold transitions. Recall that
depending on the value of $N_c g_{\text{YM}}^2$ the unflavored system
was best described by either D6-branes on the deformed conifold ($N_c
g_{\text{YM}}^2 \ll 1$) or pure two-form flux on the two-sphere in the
resolved conifold ($N_c g_{\text{YM}}^2 \gg 1$). It is a priori not
clear that this is still the case upon addition of flavors -- a
problem that could be studied using topological string theory as in
\cite{Gopakumar:1998ki}, type IIA string theory or M-theory
(\cite{Atiyah:2000zz}).

\section*{Acknowledgements}
\label{sec:acknowledgements}
We would like to thank Dario Martelli, Carlos N\'u\~nez and Ioannis
Papadimitriou for useful discussions and 
comments on the manuscript. We are also grateful
to Andreas Brandhuber, Tom\'as Ort\'in and James Sparks for further
discussions. J.S.~is supported by the German National Academic
Foundation (Studienstiftung des deutschen Volkes) and an STFC
studentship.

\appendix

\section{D6-brane embeddings}
\label{sec:brane_embeddings}
We will now discuss D6-brane embeddings in the three type IIA
reductions of the Bryant-Salamon metric
(\ref{eq:Bryant-Salamon_metric}). In principle one would have to study
each of the three reductions independently, but as we will show now it
is actually possible to search for these embeddings directly in
M-theory. Strictly speaking we will do nothing but rewriting
the calibration condition of type IIA string theory in eleven-dimensional notation. However this turns out to be quite useful, as
the M-theory expressions are more compact and less convoluted than
their lower-dimensional counterparts.

The starting point is the gauged sigma model action of
\cite{Bergshoeff:1997gy}. Here, the authors constructed an action that
is the eleven-dimensional uplift of the DBI action of a D6-brane. In
other words, it can be thought of as the world-volume action of a
Kaluza-Klein monopole. Let the M-theory circle be described by the
Killing vector $K = \partial_z$. Then
\begin{equation}
  \label{eq:kk-monopole-gauged-sigma-model}
  \begin{aligned}
    S_{\text{KK7}} &= -T_{\text{KK7}} \int \td^7\xi K^2
    \sqrt{-\det \partial_i X^M \partial_j X^N
      \Pi_{MN}} \\
    \Pi_{MN} &= g_{MN} - K^{-1} K_M K_N
  \end{aligned}
\end{equation}
The action is that of a gauged sigma
model. $\hat{\Pi}^M_{\phantom{M}N}$ projects eleven to ten-dimensional
vectors. One verifies by explicit calculation that
(\ref{eq:kk-monopole-gauged-sigma-model}) reduces to the DBI action of
a D6-brane.

We want to use $\Pi_{MN}$ to describe calibrated cycles of D6-branes in
type IIA using M-theory notation. Recall that a D6-brane embedding
$X^\mu(\xi^i)$ is supersymmetric if it satisfies the calibration
condition
\begin{equation}
  \label{eq:D6-calibration-condition}
  X^* \phi_{\text{D6}} = \sqrt{-g_{\text{ind.}}}
\end{equation}
Here $(g_{\text{ind.}})_{ij} = \partial_i X^\mu \partial_j X^\nu
g_{\mu\nu}$ is the induced metric and $\phi_{\text{D6}}$ the
calibration form
(\ref{eq:IIA_calibration_form_in_terms_of_vielbeins}). Defining
$(\hat{e}^{\text{IIA}})^A_M = \Pi_{MN} \hat{E}^{NA}$ we have, using
(\ref{eq:Scherk-Schwarz_gauge}),
$(\hat{e}^{\text{IIA}})^a = e^{-\frac{\Phi}{3}} e^a$. We can now
define the M-theory lift of the type IIA calibration form as
\begin{equation}
  \label{eq:M-theory_calibration_form_in_terms_of_vielbeins}
  \phi_{\text{KK7}} = (\hat{e}^{\text{IIA}})^{x^0x^1x^2x^3} \wedge
  \lbrack (\hat{e}^{\text{IIA}})^{125} - (\hat{e}^{\text{IIA}})^{345} -
  (\hat{e}^{\text{IIA}})^{\rho24} - (\hat{e}^{\text{IIA}})^{\rho13} \rbrack
\end{equation}
In an abuse of notation, we have labeled this the calibration form of
a KK-monopole.
Also $\sqrt{-\Pi} = e^{-\frac{7}{3}\Phi} \sqrt{-g_{\text{ind.}}}$, and we
arrive at a lifted form of the calibration condition
(\ref{eq:D6-calibration-condition}),
\begin{equation}
  \label{eq:KK7-calibration-condition}
  X^* \phi_{\text{KK7}} = \sqrt{-\Pi}
\end{equation}

We will now use (\ref{eq:KK7-calibration-condition}) to study D6-brane
embeddings. Recall that there are three $\Lie{U}(1)$ isometries, with
three different dimensional reductions
\begin{equation}
  \begin{aligned}
    \partial_{\psi+\tilde{\psi}} &\subset \sigma \times \Sigma &
    &\text{Resolved conifold} \\
    \partial_\phi &\subset \sigma & &\text{Resolved conifold} \\
    \partial_{\tilde{\phi}} &\subset \Sigma & &\text{Deformed conifold}
  \end{aligned}
\end{equation}

\paragraph{Color embeddings}
\label{sec:color-embeddings}

Color embeddings are those which wrap only a compact cycle. In the
case at hand they do not extend along the radial direction at all. If
we specify to the deformed conifold, that is, we choose the 
isometry $K = \partial_{\tilde{\phi}}$, we find an embedding
parameterized by\footnote{
The notation for these embedding diagrams is as follows: a $-$ signals
a non-compact direction along which the brane extends, a $\circ$ a
wrapped compact one. $K$ denotes the M-theory circle associated with
the Killing vector $K$, $.$ finally stands for localized
directions.}
\begin{equation}
  \begin{tabular}[c]{cccccccc}
    $x^\mu$ & $\rho$ & $\theta$ & $\phi$ & $\psi$ & $\tilde{\theta}$ &
    $\tilde{\phi}$ & $\tilde{\psi}$ \\ \hline
    $ - $ & $\rho_0$ & $\circ$ & $\circ$ & $\circ$ &
    $.$ & $K$ & $.$
  \end{tabular}
\end{equation}
The embedding exists only at $\rho = \rho_0$ as
\begin{equation}
  \begin{aligned}
    X^*\phi_{\text{KK7}} &= -\frac{2 \rho^3 + \rho_0^3}{72 \sqrt{3}}\sin\theta &
    \sqrt{-\Pi} &= \frac{-4 \rho^3 + \rho_0^3}{72 \sqrt{3}}\sin\theta
    \\
    X^*\phi_{\text{KK7}} &\overset{\rho \to \rho_0}{=} \sqrt{-\Pi}
  \end{aligned}
\end{equation}
So we recover the color brane embedding of the string dual we started
with. Note that this cycle is calibrated in M-theory though. I.e.~it
is a minimum volume cycle of the eleven-dimensional geometry.

For the resolved conifold associated with $K = \partial_{\phi}$ one
might try an embedding as
\begin{equation}
  \begin{tabular}[c]{cccccccc}
    $x^\mu$ & $\rho$ & $\theta$ & $\phi$ & $\psi$ & $\tilde{\theta}$ &
    $\tilde{\phi}$ & $\tilde{\psi}$ \\ \hline
    $ - $ & $\rho_0$ & $.$ & $K$ & $.$ & $\circ$ & $\circ$ & $\circ$
  \end{tabular}
\end{equation}
However, the cycle in question vanishes at $\rho = \rho_0$, as one
would expect.

\paragraph{Massless flavor embeddings}
\label{sec:massl-flav-embedd}

Massless flavor branes extend fully along the radial direction
$\rho$. Therefore they only need to wrap a two-cycle in the internal
geometry and one can make the following guess
\begin{equation}
  \begin{tabular}[c]{cccccccc}
    $x^\mu$ & $\rho$ & $\theta$ & $\phi$ & $\psi$ & $\tilde{\theta}$ &
    $\tilde{\phi}$ & $\tilde{\psi}$ \\ \hline
    $ - $ & $-$ & $.$ & $.$ & $\circ$ & $.$ & $.$ & $\circ$
  \end{tabular}
\end{equation}
Note that this embedding works for both the $\partial_\phi$ and the
$\partial_{\tilde{\phi}}$ isometries.

For the deformed conifold, i.e.~reduction along
$\partial_{\tilde{\phi}}$, we obtain the relation
\begin{equation}
  \begin{aligned}
    X^*\phi_{\text{KK7}} &= \frac{\rho^2}{6\sqrt{3}}
    \sin^2\tilde{\theta} \\ 
    \sqrt{-\Pi} &= \frac{\rho^2}{6\sqrt{3}} \sin\tilde{\theta}
  \end{aligned}
\end{equation}
demanding $\tilde{\theta} = \frac{\pi}{2}$. The resolved conifold
associated with $\partial_\phi$ gives
\begin{equation}
  \begin{aligned}
    X^*\phi_{\text{KK7}} &= \frac{\rho^2}{6\sqrt{3}} \sin^2\theta \\
    \sqrt{-\Pi} &= \frac{\rho^2}{6\sqrt{3}} \sin\theta
  \end{aligned}
\end{equation}
demanding $\theta = \frac{\pi}{2}$; whereas for the $\partial_\psi
+ \partial_{\tilde{\psi}}$ reduction both $X^*\phi$ and $\sqrt{-\Pi}$
  vanish.
Interestingly, in M-theory the cycle $(x^\mu, \rho, \psi,
\tilde{\psi})$ is calibrated in the traditional sense; that is, it is
a minimal volume cycle.

\paragraph{Massive flavor embeddings}
\label{sec:massive-flav-embedd}
Naturally one would like to relax the constraints on $\theta$ and
$\tilde{\theta}$ respectively from the above paragraph. A good guess
to do so lies in assuming a relation between $\rho$ and $\theta$ (or
$\tilde{\theta}$).

In the case of the $\partial_{\tilde{\phi}}$ reduction, we assume
\begin{equation}
  \begin{tabular}[c]{cccccccc}
    $x^\mu$ & $\rho$ & $\theta$ & $\phi$ & $\psi$ & $\tilde{\theta}$ &
    $\tilde{\phi}$ & $\tilde{\psi}$ \\ \hline
    $ - $ & $\rho(\tilde{\theta})$ & $.$ & $.$ & $\circ$ & $\circ$ & $.$ &
    $\circ$
  \end{tabular}
\end{equation}
Then
\begin{equation}
  \begin{aligned}
    X^*\phi_{\text{KK7}} &= \frac{(\rho^3-\rho_0^3)\cos\tilde{\theta}
      + 3 \rho^2 \rho^\prime \sin\tilde{\theta}}{18\sqrt{3}}
    \sin\tilde{\theta} \\
    \sqrt{-\Pi} &= \frac{\sqrt{(\rho^3 - \rho_0^3)^2 + 9 \rho^4
        (\rho^\prime)^2 }}{18\sqrt{3}} \sin\tilde{\theta}
  \end{aligned}
\end{equation}
Demanding the two expressions to agree, it follows that
\begin{equation}
  \begin{aligned}
    \rho^\prime(\tilde{\theta}) &= \frac{\rho^3 - \rho_0^3}{3 \rho^2}
    \tan\tilde{\theta} \\
    \rho(\tilde{\theta}) &= \left( \rho_0^3 + e^{3 C_1}
      \sec\tilde{\theta} \right)^{1/3}
  \end{aligned}
\end{equation}
with $C_1$ being a constant of integration, associated with the mass
of the flavors, as we will show now. $\sec\tilde{\theta} \in \lbrack
1, \infty )$, so the brane reaches down to $(\rho_0^3 + e^{3
  C_1})^{1/3}$. Thus the massless limit is given by $C_1 \to
-\infty$. In order to compare this embedding with the massless one of
the previous paragraph, we have to solve the embedding equation for
$\tilde{\theta}$ before taking this limit -- as we expect the brane to
be localized in $\tilde{\theta}$, so the mapping $\tilde{\theta}
\mapsto \rho$ is ill defined. The result is
\begin{equation}
  \begin{aligned}\label{eq:massive_embedding}
    \tilde{\theta} &= \arccos \frac{e^{3 C_1}}{\rho^3 - \rho_0^3}
  \end{aligned}
\end{equation}
So in the limit $C_1 \to -\infty$, the brane sits once more at
$\tilde{\theta} = \frac{\pi}{2}$, which is also the position of the
brane for $\rho \gg \rho_0$.

For the $\partial_\phi$ reduction, one needs to swap $\theta$ for
$\tilde{\theta}$. Then, 
\begin{equation}
  \begin{tabular}[c]{cccccccc}
    $x^\mu$ & $\rho$ & $\theta$ & $\phi$ & $\psi$ & $\tilde{\theta}$ &
    $\tilde{\phi}$ & $\tilde{\psi}$ \\ \hline
    $ - $ & $\rho(\theta)$ & $\circ$ & $.$ & $\circ$ & $.$ & $.$ &
    $\circ$
  \end{tabular}
\end{equation}
The calibration condition is given by
\begin{equation}
  \begin{aligned}
    X^*\phi_{\text{KK7}} &= \frac{(8\rho^6 - 7 \rho^3 \rho_0^3 - \rho_0^6)
      \cos\theta + 6\rho^2 (4\rho^3 - \rho_0^3) \rho^\prime
      \sin\theta}{36 \sqrt{3} (4\rho^3 - \rho_0^3)} \sin\theta \\
    \sqrt{-\Pi} &= \frac{\sqrt{4\rho^6 - 6\rho^3 \rho_0^3 + \rho_0^6 +
        36 \rho^4 (\rho^\prime)^2 }}{36 \sqrt{3}} \sin\theta
  \end{aligned}
\end{equation}
leading to a differential equation for $\rho$ that is considerably
harder to solve than the previous one. One can study it numerically,
obtaining results similar to those of the previous embedding. As to
analytic results, setting $\rho_0 \to 0$, leads to simplifications
allowing for
\begin{equation}
  \rho(\theta) = C_1 (\sec\theta)^{1/3}  
\end{equation}
which is identical to~(\ref{eq:massive_embedding}) in the same limit.

\section{Kaluza-Klein reduction of supergravity variations with torsion}
\label{sec:kk-reduction_with_torsion}

We review the dimensional reduction of the SUSY variations
-- with an additional torsion term -- from eleven to ten-dimensional
supergravity. Conceptually we follow \cite{Ortin:2004ms}, our
conventions are slightly different though. We assume a space-time
with topology $\mathcal{M}_{10}\times S^1$ and label the eleventh coordinate as
$z$. Naturally all fields will be independent of $z$. Further assuming
the eleven-dimensional background to be purely gravitational, we only need to
consider the variation of the gravitino,
\begin{equation}
  \label{eq:dim-red_susy-trans_11d-gravition}
  \delta_{\hat \epsilon} \hat{\psi}_M = \partial_M \hat{\epsilon} +
  \frac{1}{4} \hat{\omega}_{MAB} \Gamma^{AB} \hat{\epsilon} +
  \frac{1}{4} \hat{\tau}_{MAB} \hat{\Gamma}^{AB} \hat{\epsilon}
\end{equation}
which we have modified by the presence of the torsion term
$\hat{\tau}$. As in section \ref{sec:brandhubers-paper} we take the
vielbein to be in Scherk-Schwarz gauge
(\ref{eq:Scherk-Schwarz_gauge}).

We shall perform the reduction of
\eqref{eq:dim-red_susy-trans_11d-gravition} step by step and our first
aim shall be the reduction of the spin connection
\begin{equation}
  \label{eq:spin-connection-in-Omegas}
  \hat{\omega}_{ABC} = \frac{1}{2} \left( \hat{\Omega}_{CAB} -
    \hat{\Omega}_{BAC} - \hat{\Omega}_{ABC} \right)
\end{equation}
with the objects of anholomorphicity defined as
\begin{equation}
  \label{eq:spin-connection-Omegas}
  \hat{\Omega}_{ABC} = \left( \partial_M \hat{e}_N^K
    - \partial_N \hat{e}_M^K \right) \hat{\eta}_{KA} \hat{E}^N_B
  \hat{E}^M_C
\end{equation}
Then
\begin{equation}
  \label{eq:omegas}
  \begin{aligned}
    \hat{\omega}_{zbc} &= + \frac{e^{\frac{4}{3}\Phi}}{2} (\td A)_{bc} &
    \hat{\omega}_{abc} &= \frac{e^{\frac{1}{3}\Phi}}{3} \left( 
      \eta_{ab} \partial_c \Phi - \eta_{ac} \partial_b \Phi
    \right) + e^{\frac{1}{3}\Phi} \omega_{abc} \\
    \hat{\omega}_{abz} &= -\frac{e^{\frac{4}{3}\Phi}}{2} (\td A)_{ab} &
    \hat{\omega}_{zaz} &= \frac{2}{3} e^{\frac{1}{3}\Phi} \partial_a
    \Phi
  \end{aligned}
\end{equation}
Note that we use $\td A_{\mu\nu}$ instead of $F_{\mu\nu}$ as we are
anticipating the inclusion of sources such that $F$ is no longer exact.

Turning to the gravitino, one could make an ansatz
\begin{equation}
  \label{eq:gravitino_reduction_first_ansatz}
  \hat{\psi}_M = \left( e^{m\Phi} \psi_\mu, e^{n\Phi} \lambda \right)
\end{equation}
and
\begin{equation}
  \label{eq:susy-spinor_reduction_first_ansatz}
  \hat{\epsilon} = e^{l\Phi} \epsilon
\end{equation}
with $l,m,n \in \mathbb{C}$. Yet, as we will see, we will need to
consider linear combinations such as $\hat{\psi}_\mu =
e^{m\Phi}\psi_\mu + e^{n\Phi}\Gamma_\mu \lambda + e^{p\Phi}\Gamma_\mu
\Gamma^{11} \lambda$.

We begin with the covariant derivative of the SUSY spinor, looking
first at the vector components:
\begin{equation}
  \label{eq:susy-spinor_reduction_vector-component}
  \begin{aligned}
    e^{-l\Phi} \hat{D}_\mu \hat{\epsilon} &= 
    ( l \partial_\mu \Phi \epsilon + \partial_\mu \epsilon )
    + e_\mu^a \left\lbrack \frac{1}{4} \omega_{abc} +
      \frac{1}{12} ( \eta_{ab} \partial_c \Phi - \eta_{ac} \partial_b
      \Phi ) \right\rbrack \Gamma^{bc} \epsilon \\
    &- \frac{1}{4} e^\Phi e_\mu^a \td A_{ab} \Gamma^b \Gamma^{11} \epsilon
    +\left( \frac{1}{8} e^{2\Phi} A_\mu \td A_{bc} \Gamma^{bc} +
      \frac{1}{3} e^\Phi A_\mu \partial_b \Phi \Gamma^b \Gamma^{11}
    \right) \epsilon \\
    &+ \frac{1}{4} \hat{\tau}_{\mu bc} \Gamma^{bc} \epsilon
    + \frac{1}{2} \hat{\tau}_{\mu bz} \Gamma^b \Gamma^{11} \epsilon
  \end{aligned}
\end{equation}
The scalar component satisfies
\begin{equation}
  \label{eq:susy-spinor_reduction_scalar-component}
  \begin{aligned}
    e^{-l\Phi} \hat{D}_z \hat{\epsilon} 
    &= \frac{e^{2\Phi}}{8} \td A_{bc} \Gamma^{bc} \epsilon
    + \frac{e^\Phi}{3} \partial_b \Phi \Gamma^b
    \Gamma^{11} \epsilon
    + \frac{1}{4} \hat{\tau}_{zbc} \Gamma^{bc} \epsilon + \frac{1}{2}
    \hat{\tau}_{zbz} \Gamma^b \Gamma^{11} \epsilon
  \end{aligned}
\end{equation}

Equations (\ref{eq:susy-spinor_reduction_vector-component}) and
(\ref{eq:susy-spinor_reduction_scalar-component}) hold in string
frame. To convert to Einstein frame we need to recall that the gamma
matrices are defined in tangent space, from which it follows that only
the curved space gamma matrices are affected by Weyl
transformations. For a generic Weyl transformation, we have
\begin{equation}
  \label{eq:dilaton-frame-conversion}
  \begin{aligned}
    e_\mu^a &\mapsto e^{\delta \Phi} e_\mu^a &
    \Omega_{abc} &\mapsto e^{-\delta \Phi} \Omega_{abc} +
    e^{-\delta \Phi} \delta ( \eta_{ab} \partial_c \Phi -
    \eta_{ac} \partial_b \Phi ) \\
    E_a^\mu &\mapsto e^{-\delta \Phi} E_a^\mu &
    \omega_{abc} &\mapsto e^{-\delta \Phi} \omega_{abc} - \delta
    e^{-\delta \Phi} ( \eta_{ab} \partial_c \Phi - \eta_{ac} \partial_b
    \Phi ) \\
    \partial_a &\mapsto e^{-\delta \Phi} \partial_a &
    \td A_{a_1 \dots a_p} &\mapsto e^{-p\delta \Phi} \td A_{a_1 \dots a_p} \\
    \Gamma^a &\mapsto \Gamma^a &
    \eta_{ab} &\mapsto \eta_{ab} \\
    \hat{\tau}_{\mu bc} &\mapsto \hat{\tau}_{\mu bc}
  \end{aligned}
\end{equation}
So that with $(e^S)_\mu^a =
e^{\frac{1}{4}\Phi} (e^E)_\mu^a$, $\delta = \frac{1}{4}$,
\begin{equation}
  \label{eq:susy-spinor_reduction_Einstein-frame}
  \begin{aligned}
    e^{-l\Phi} \hat{D}_\mu \hat{\epsilon} 
    &= \left( l \partial_\mu
      \Phi \epsilon + \partial_\mu \epsilon \right)
    + e_\mu^a \left\lbrack \frac{1}{4} \omega_{abc} +
      \frac{1}{48} ( \eta_{ab} \partial_c \Phi - \eta_{ac} \partial_b
      \Phi ) \right\rbrack \Gamma^{bc} \epsilon \\
    &- \frac{1}{4} e^{\frac{3}{4}\Phi} e_\mu^a \td A_{ab} \Gamma^b
    \Gamma^{11} \epsilon
    +\left( \frac{1}{8} e^{\frac{3}{2}\Phi} A_\mu \td A_{bc} \Gamma^{bc}
      + \frac{1}{3} e^{\frac{3}{4}\Phi} A_\mu \partial_b \Phi \Gamma^b
      \Gamma^{11} \right) \epsilon \\
    &+ \frac{1}{4} \hat{\tau}_{\mu bc} \Gamma^{bc} \epsilon
    + \frac{1}{2} \hat{\tau}_{\mu bz} \Gamma^b \Gamma^{11} \epsilon \\
    e^{-l\Phi} \hat{D}_z \hat{\epsilon} &=
    \frac{e^{\frac{3}{2}\Phi}}{8} \td A_{bc} \Gamma^{bc} \epsilon +
    \frac{e^{\frac{3}{4}\Phi}}{3} \partial_b \Phi \Gamma^b
    \Gamma^{11} \epsilon
    + \frac{1}{4} \hat{\tau}_{zbc} \Gamma^{bc} \epsilon + \frac{1}{2}
    \hat{\tau}_{zbz} \Gamma^b \Gamma^{11} \epsilon
  \end{aligned}
\end{equation}

One needs to compare (\ref{eq:susy-spinor_reduction_vector-component})
and (\ref{eq:susy-spinor_reduction_scalar-component}) or
(\ref{eq:susy-spinor_reduction_Einstein-frame}) repectively to the
SUSY variations of the ansatz
(\ref{eq:gravitino_reduction_first_ansatz})
\begin{equation}
  \label{eq:susy-spinor_reduction_susy-variation}
  \begin{aligned}
    \hat{D}_\mu \hat{\epsilon} &= \delta_{\hat{\epsilon}}
    \hat{\psi}_\mu = e^{m\Phi} \left( m \delta_{\hat{\epsilon}} \Phi
      \psi_\mu + \delta_{\hat{\epsilon}} \psi_\mu \right) = e^{m\Phi}
    \delta_{\hat{\epsilon}} \psi_\mu \\
    \hat{D}_z \hat{\epsilon} &= \delta_{\hat{\epsilon}} \hat{\psi}_z =
    e^{n\Phi} \left( n \delta_{\hat{\epsilon}} \Phi \lambda +
      \delta_{\hat{\epsilon}} \lambda \right) = e^{n\Phi}
    \delta_{\hat{\epsilon}} \lambda 
  \end{aligned}
\end{equation}
The last equalities follow from the fact that we assume the spinor
fields to vanish. However the resulting variations will explicitly
depend on the gauge-potential $A$. We therefore replace the
original ansatz (\ref{eq:gravitino_reduction_first_ansatz}) with
\begin{equation}
  \label{eq:gravitino_reduction_final_ansatz}
  \begin{aligned}
    \psi_\mu &= \hat{\psi}_\mu - x_2 e_\mu^a \eta_{ab} \Gamma^b
    \Gamma^{11} \hat{\psi}_z - x_3 A_\mu \hat{\psi}_z \\
    \lambda &= x_1 \hat{\psi}_z \\
    \hat{\epsilon} &= e^{l\Phi} \epsilon
  \end{aligned}
\end{equation}
which amounts to a field redefinition in ten dimensions. If one was to
work properly, one had to peform the dimensional reduction of the
action as well in order to make sure that the fermion terms have the
proper normalizations. The SUSY variations of
(\ref{eq:gravitino_reduction_final_ansatz}) are
\begin{equation}
\label{eq:ten_dim_susy_transforms_from_eleven_dim_ones}
  \begin{aligned}
    \delta_\epsilon \psi_\mu &= \delta_\epsilon \hat{\psi}_\mu - x_2
    e_\mu^a \eta_{ab} \Gamma^b \Gamma^{11} \delta_\epsilon
    \hat{\psi}_z - x_3 A_\mu \delta_\epsilon \hat{\psi}_z \\
    &= e^{-l\Phi} \hat{D}_\mu \hat{\epsilon} - x_2 e_\mu^a \eta_{ab}
    \Gamma^b \Gamma^{11} e^{-l\Phi} \hat{D}_z \hat{\epsilon} - x_3
    A_\mu e^{-l\Phi} \hat{D}_z \hat{\epsilon} \\
    \delta_\epsilon \lambda &= x_1 e^{-l\Phi} \hat{D}_z
    \hat{\epsilon}
  \end{aligned}
\end{equation}
Note that the variations of the bosonic fields all vanish, as we have
set the fermions explicitly to zero. Our aim is to compare
(\ref{eq:ten_dim_susy_transforms_from_eleven_dim_ones}) with the
IIA Einstein frame SUSY variations as taken from
\cite{Paredes:2004xw}
\begin{subequations} \label{eq:IIA_susy-variations_EinsteinFrame}
  \begin{align}
    \delta \lambda &= \frac{\sqrt{2}}{4} \partial_\mu \Phi \Gamma^\mu
    \Gamma^{11} \epsilon + \frac{3}{16} \frac{1}{\sqrt{2}}
      e^{\frac{3}{4}\Phi} \td A_{\mu_1 \mu_2} \Gamma^{\mu_1 \mu_2}
      \epsilon \label{eq:IIA_dilatino-variation_EinsteinFrame}\\
    \delta \psi_\mu &= D_\mu \epsilon + \frac{1}{64}
    e^{\frac{3}{4}\Phi} \td A_{\mu_1 \mu_2} \left(
      \Gamma_\mu^{\phantom{\mu} \mu_1 \mu_2} - 14 \delta_{\mu}^{\mu_1}
      \Gamma^{\mu_2} \right) \Gamma^{11}
    \epsilon \label{eq:IIA_gravitino-variation_EinsteinFrame}
  \end{align}
\end{subequations}

Before evaluating
(\ref{eq:ten_dim_susy_transforms_from_eleven_dim_ones}), we calculate\footnote{
The following is used here:
\begin{equation*}
  \begin{aligned}
    \Gamma^a \Gamma^b &= \Gamma^{ab} + \eta^{ab} \\
    \Gamma^a \Gamma^b \Gamma^c &= \Gamma^{abc} + \eta^{ab} \Gamma^c -
    \eta^{ca} \Gamma^b + \eta^{bc} \Gamma^a
  \end{aligned}
\end{equation*}
}
\begin{equation}\label{eq:susy_spinor_reduction_field_redefinition_helper}
  \begin{aligned}
    &x_2 e_\mu^a \eta_{ab} \Gamma^b \Gamma^{11} e^{-l\Phi} \hat{D}_z
    \hat{\epsilon} \\
    &= x_2 \frac{1}{8} e^{\frac{3}{2}\Phi} e_\mu^a
    \eta_{ab} \td A_{cd} ( \Gamma^{bcd} + 2\eta^{bc} \Gamma^d )
    \Gamma^{11} \epsilon \\
    &- x_2 \frac{1}{3} e^{\frac{3}{4}\Phi} e_\mu^a \partial_a \Phi
    \epsilon
    - x_2 \frac{1}{6} e^{\frac{3}{4}\Phi} e_\mu^a
    (\eta_{ab} \partial_c \Phi - \eta_{ac} \partial_b \Phi)
    \Gamma^{bc}\epsilon \\
    &- \frac{1}{2} x_2 e_\mu^a \eta_{ab} \hat{\tau}_{zaz} \epsilon
    -\frac{1}{2} x_2 e_\mu^a \eta_{ab} \hat{\tau}_{zcz} \Gamma^{bc}
    \epsilon
    + \frac{1}{4} x_2 e_\mu^a \hat{\tau}_{zcd} (\eta_{ab} \Gamma^{bcd} +
    2 \delta_a^c \Gamma^d) \Gamma^{11} \epsilon
  \end{aligned}
\end{equation}
Putting things together, we use equations
(\ref{eq:susy-spinor_reduction_Einstein-frame}) and
(\ref{eq:ten_dim_susy_transforms_from_eleven_dim_ones})
\begin{equation}\label{eq:susy_spinor_reductions-the_BIG_calculation}
  \begin{aligned}
    \delta_\epsilon \psi_\mu &= \left( l \partial_\mu
      \Phi \epsilon + \partial_\mu \epsilon \right)
    + e_\mu^a \left\lbrack \frac{1}{4} \omega_{abc} +
      \frac{1}{48} ( \eta_{ab} \partial_c \Phi - \eta_{ac} \partial_b
      \Phi ) \right\rbrack \Gamma^{bc} \epsilon \\
    &- \frac{1}{4} e^{\frac{3}{4}\Phi} e_\mu^a \td A_{ab} \Gamma^b
    \Gamma^{11} \epsilon \\
    &-x_2 \frac{1}{8} e^{\frac{3}{2}\Phi} e_\mu
    \eta_{ab} \td A_{cd} ( \Gamma^{bcd} + 2\eta^{bc} \Gamma^d )
    \Gamma^{11} \epsilon \\
    &+ x_2 \frac{1}{3} e^{\frac{3}{4}\Phi} e_\mu^a \partial_a \Phi
    \epsilon + x_2 \frac{1}{6} e^{\frac{3}{4}\Phi} e_\mu^a
    (\eta_{ab} \partial_c \Phi - \eta_{ac} \partial_b \Phi)
    \Gamma^{bc}\epsilon \\
    &+\left( \frac{1}{8} e^{\frac{3}{2}\Phi} A_\mu \td A_{bc} \Gamma^{bc}
      + \frac{1}{3} e^{\frac{3}{4}\Phi} A_\mu \partial_b \Phi \Gamma^b
      \Gamma^{11} \right) \epsilon \\
    &-x_3 \left( \frac{e^{\frac{3}{2}\Phi}}{8} e_\mu^a A_a \td A_{bc} \Gamma^{bc}
      + \frac{e^{\frac{3}{4}\Phi}}{3} e_\mu^a A_a \partial_b \Phi
      \Gamma^b \Gamma^{11} \right) \epsilon \\
    &+ \frac{1}{4} \hat{\tau}_{\mu bc} \Gamma^{bc} \epsilon
    + \frac{1}{2} \hat{\tau}_{\mu bz} \Gamma^b \Gamma^{11} \epsilon \\
    &+ \frac{1}{2} x_2 e_\mu^a \eta_{ab} \hat{\tau}_{zaz} \epsilon
    +\frac{1}{2} x_2 e_\mu^a \eta_{ab} \hat{\tau}_{zcz} \Gamma^{bc}
    \epsilon
    - \frac{1}{4} x_2 e_\mu^a \hat{\tau}_{zcd} (\eta_{ab} \Gamma^{bcd} +
    2 \delta_a^c \Gamma^d) \Gamma^{11} \epsilon \\
    &- x_3 A_\mu (\frac{1}{4} \hat{\tau}_{zbc} \Gamma^{bc} + \frac{1}{2}
    \hat{\tau}_{zbz} \Gamma^b \Gamma^{11} )\epsilon \\
    \delta_\epsilon \lambda &=
    x_1 \frac{e^{\frac{3}{2}\Phi}}{8} \td A_{bc} \Gamma^{bc} \epsilon +
    x_1 \frac{e^{\frac{3}{4}\Phi}}{3} \partial_b \Phi \Gamma^b
    \Gamma^{11} \epsilon
    + x_1 (\frac{1}{4} \hat{\tau}_{zbc} \Gamma^{bc}
    + \frac{1}{2} \hat{\tau}_{zbz} \Gamma^b \Gamma^{11} )\epsilon
  \end{aligned}
\end{equation}
Investigating this and comparing with
(\ref{eq:IIA_susy-variations_EinsteinFrame}) one sets $l =
\frac{1}{24}$ and
\begin{equation}\label{eq:susy_spinor_reduction_parameters}
  \begin{aligned}
    x_1 &= \frac{3\sqrt{2}}{4} e^{-\frac{3}{4}\Phi} \\
    x_2 &= -\frac{1}{8}e^{-\frac{3}{4}\Phi} \\
    x_3 &= 1
  \end{aligned}
\end{equation}
to obtain the standard type IIA SUSY variations garnished with some
additional torsion terms:
\begin{equation}\label{eq:susy_spinor_reduction_final_result}
  \begin{aligned}
    \delta_\epsilon \psi_\mu &= \partial_\mu \epsilon  
    + e_\mu^a \frac{1}{4} \omega_{abc} \Gamma^{bc} \epsilon
    + \frac{1}{64} e^{\frac{3}{4}\Phi} e_\mu^a \td A_{cd} \left(
      \eta_{ab} \Gamma^{bcd} - 14 \delta_a^c \Gamma^d \right)
    \Gamma^{11} \epsilon \\
    &+ \frac{1}{4} \hat{\tau}_{\mu bc} \Gamma^{bc} \epsilon
    + \frac{1}{2} \hat{\tau}_{\mu bz} \Gamma^b \Gamma^{11} \epsilon \\
    &- \frac{1}{16} e^{-\frac{3}{4}\Phi} e_\mu^a \eta_{ab}
    \hat{\tau}_{zaz} \epsilon 
    -\frac{1}{16} e^{-\frac{3}{4}\Phi} e_\mu^a \eta_{ab} \hat{\tau}_{zcz}
    \Gamma^{bc} \epsilon \\
    &+ \frac{1}{32} e^{-\frac{3}{4}\Phi} e_\mu^a \hat{\tau}_{zcd}
    (\eta_{ab} \Gamma^{bcd} +
    2 \delta_a^c \Gamma^d) \Gamma^{11} \epsilon \\
    &- A_\mu (\frac{1}{4} \hat{\tau}_{zbc} \Gamma^{bc} + \frac{1}{2}
    \hat{\tau}_{zbz} \Gamma^b \Gamma^{11} )\epsilon \\
    \delta_\epsilon \lambda &=
    \frac{3}{16} \frac{1}{\sqrt{2}} e^{\frac{3}{4}\Phi} ( \td A_{bc}
    + 2 e^{-\frac{3}{2}\Phi} \hat{\tau}_{zbc} ) \Gamma^{bc}
    \epsilon + \frac{\sqrt{2}}{4} (\partial_b \Phi +  \frac{3}{2}
    e^{-\frac{3}{4}\Phi} \hat{\tau}_{zbz} )\Gamma^b
    \Gamma^{11}\epsilon
  \end{aligned}
\end{equation}

\bibliographystyle{plain}

\end{document}